\documentclass[a4paper, 10pt, conference]{ieeeconf}      

\IEEEoverridecommandlockouts                              
\makeatletter
\let\NAT@parse\undefined
\makeatother

\usepackage{cite}
\usepackage{amsmath,amssymb,amsfonts}
\usepackage{algorithmic}
\usepackage{graphicx}
\usepackage{textcomp}
\usepackage{soul}
\soulregister\ref7
\soulregister\cite7

\usepackage{makecell}
\usepackage{lineno}
\modulolinenumbers[5]
\usepackage{subfigure}
\usepackage{multirow}
\let\labelindent\relax
\usepackage{enumitem}
\usepackage{colortbl}
\usepackage{url}

\definecolor{mygray}{gray}{.9}

\def\BibTeX{{\rm B\kern-.05em{\sc i\kern-.025em b}\kern-.08em
		T\kern-.1667em\lower.7ex\hbox{E}\kern-.125emX}}

\title{\LARGE \bf
Improving Recommendation Diversity by Highlighting the ExTrA Fabricated Experts
}

\author{Ya-Hui An$^{1}$ and Qiang Dong$^{1}$ and Quan Yuan$^{1}$ and Chao Wang$^{2}$
\thanks{$^{1}$CompleX Lab, School of Computer Science and Engineering, University of Electronic Science and Technology of China, Chengdu 611731, China}%
\thanks{$^{2}$School of Electrical Engineering, Chongqing University, Chongqing 400044, China}%
}

\begin{document}

\maketitle
\thispagestyle{empty}
\pagestyle{empty}

\begin{abstract}
	\noindent
	Nowadays, recommender systems (RSes) are becoming increasingly important to individual users and business marketing, especially in the online e-commerce scenarios. However, while the majority of recommendation algorithms proposed in the literature have focused their efforts on improving prediction accuracy, other important aspects of recommendation quality, such as diversity of recommendations, have been more or less overlooked. In the latest decade, recommendation diversity has drawn more research attention, especially in the models based on user-item bipartite networks. In this paper, we introduce a family of approaches to extract fabricated experts from users in RSes, named as the Expert Tracking Approaches (ExTrA for short), and explore the capability of these fabricated experts in improving the recommendation diversity, by highlighting them in a well-known bipartite network-based method, called the Mass Diffusion (MD for short) model. These ExTrA-based models are compared with two state-of-the-art MD-improved models HHP and BHC, with respect to recommendation accuracy and diversity. Comprehensive empirical results on three real-world datasets MovieLens, Netflix and RYM show that, our proposed ExTrA-based models can achieve significant diversity gain while maintain comparable level of recommendation accuracy. 
\end{abstract}

\section{Introduction}
\label{sec:introduction}
Recommender systems (RSes) are powerful tools of helping users confront the challenge of information overload, by uncovering users' potential preferences on uncollected items and accordingly delivering personalized recommendation lists. Accuracy used to be regarded as the most important concern for RSes~\cite{koren2008factorization,sarwar2001item,CaoDa2018AGR,XiangnanHe2017BTRo,ZhangFuzheng2016CKBE,WangHongwei2018RPUP}. However, with the fast development of on-line e-commercial services, users' satisfaction with RSes is not only related to recommendation accuracy, but also dependent on the diversity, which measures the personalization levels of recommendation results~\cite{mcnee2006being,BrynjolfssonErik2006Fntr,fleder2009blockbuster,gollapudi2009axiomatic,zhou2010solving,hurley2011novelty,ashkan2015optimal,belem2016beyond,nguyen2018user}. 
However, people found that accuracy and diversity seem to be two sides of the seesaw: when one side rises, the other side falls~\cite{zhou2010solving}. Recommending more popular items would result in high accuracy but low diversity, while recommending more niches would bring high diversity but low accuracy. Diffusion-based recommendation is a vital branch to solve this accuracy-diversity dilemma in recommender systems, which makes recommendations for users by simulating a basic physical dynamic process on the user-item bipartite network ~\cite{zhou2007bipartite,zhou2010solving,an2016diffusion,nie2015information,lu2011information,liu2011information,zeng2014uncovering}.  The Mass Diffusion (MD)~\cite{zhou2007bipartite} model is the pioneer of diffusion-based recommendation methods, which works as follows. Initially, each item collected by the target user is assigned one-unit resource. Then the resource is redistributed among all the items through a two-step allocation process on the user-item bipartite graph, first from each item averagely to its neighbor users, then from each user averagely to its neighbor items. MD can achieve more accurate recommendations than the traditional item-based Collaborative Filtering (CF) model \cite{zhou2007bipartite}, although it can be categorized into a special case of CF with the RA similarity rather than the common Cosine or Jaccard similarity \cite{YuFei2016NraA}. Another model, called the Heat Conduction (HC)~\cite{zhang2007heat}, is a similar process, but allocating resource in a different way, which results in the exposure of more niches, however with rather low accuracy, thus, could not be applied alone in real RSes. Subsequently, Liu et al.~\cite{liu2011information} proposed a biased heat conduction (BHC) model to enhance the accuracy of HC model; Zhou et al.~\cite{zhou2010solving} integrated MD and HC methods together to generate a hybrid recommendation model, which improves accuracy and diversity simultaneously. 

Clearly, a system producing more personalized results could satisfy different kinds of users, and meanwhile, facilitate the huge niche market. However, the difficulty therein lies in the lack of enough usage data of niche items for RSes to mine from. In this so-called cold-start scenarios, state-of-the-art HHP and BHC models will lose their performance advantages on accuracy and diversity as illustrated in Section \ref{sec:analysis}. A straightforward way to solve this problem is to involve more side information, for example, ~\cite{wang2014improving} developed a content-based model that automatically extracts features from audio content, ~\cite{forsati2014matrix} exploited explicit trust and distrust (social) side information, and ~\cite{xie2016learning} combined matrix factorization with side information for click prediction of web advertisements. 

However, there are many restrictions in the real application scenarios when accessing and utilizing side information. For example, it's hard to acquire useful side information, and also, adding more side information aggravates the inefficiency of the RSes. Thus, in the point of view of practical applications for large scale online services, we propose to simply modify the first resource allocation step of the MD model, by assigning more resource to the fabricated expert users, instead of averagely to all the neighbor users of an item. These fabricated expert users are expected to have better capability to help the target user find relevant and diverse items. Then, the solution to the cold-start problem reduces to the approaches of extracting fabricated experts from users in the systems, dubbed as the \emph{\textbf{\underline{Ex}}pert \textbf{\underline{Tr}}acking \textbf{\underline{A}}pproaches}, (\textit{ExTrA} for short), and the corresponding fabricated expert users are called ExTrA users, ExTrA experts, or just experts for short.

This paper focuses on improving the diversity with no or trivial accuracy loss. The straightforward candidates for ExTrA users are the highly-active users who collected many items, because they are good at discovering both popular and niche items. Yet, the long-tail phenomenon exists in the active levels of users, which means that most of the users are not that active in the system. Although the highly-active users will be more good at exploring both popular and niche items, it doesn't mean that the low-active users could not help to achieve this goal. Let's take the movie watching records of two persons as an example. John is a low-active user who collected only 3 movies: \emph{Green Book(2018)}, \emph{Jaws(1975)}, and \emph{The Lobster(2015)}. Green Book is a 2018 Oscar movie, which is very famous and popular. Jaws is very famous, however, whose popularity have decayed over time in the  time-aware data set. The Lobster has never been popular (in a generalized concept of popularity). Mary is a highly-active user who collected 1000 movies, which are all famous ones. Recommending Mary's selections to most users would improve the accuracy of the system, but John can help people who have the same niche interests to find the wanted movie (The Lobster). Thus, the users who have collected many diverse items should also be recruited into the ExTrA users, and more resource should be assigned to these users in the first step of diffusion process of the MD model.

In this paper, our main contributions are three folded: 
\begin{itemize}
	\item We propose a family of fabricated expert extraction methods inspired by different intuitions, highlighting these experts may be helpful to improve the performance of many existing recommendation models.
	
	\item Comprehensive empirical results show that the ExTrA-based methods can achieve significant diversity improvement, while the recommendation accuracy is comparable with state-of-the-art HHP and BHC models.
	
	\item Our contribution is not proposing a better expert extraction approach for more accurate predictions, but aiming at highlighting the significance of the concept of experts in improving recommendation diversity of RSes.
	
\end{itemize}

\section{Framework of ExTrA-based models}
\label{key}
In this section, we first introduce the standard diffusion-based method, the MD model. Then, we present how to incorporate the ExTrA experts into the MD model, called ExTrA-based model. Finally, we tentatively explore what kinds of fabricated experts might help to improve the diversity when applied to the MD model.

\subsection{The mass diffusion model}

In this paper, we use $u$,$v$ to denote users, and $i$, $j$ items. Let $A$ be a user-item matrix, where the value of element $a_{{u}{i}}$ in $A$ represents whether user $u$ has collected item $i$ ($a_{ui} = 1$) or not ($a_{ui}= 0)$. Let  $U$ and $I$ represent the user and item sets respectively. For user $u$, we denote his/her active level (or degree) as $k_{u}$ ($u$ has collected $k_{u}$ items) and the popularity (or degree) of item $i$ as $k_{i}$ ($i$ has been collected by $k_{i}$ users). 
The user-item matrix can also be represented by a bipartite network, in which users and items are represented by nodes, user $u$ and item $i$ are connected by an edge iff the value of $a_{u i}$ is 1.

\begin{figure}
	\centering
	\includegraphics[width=\columnwidth]{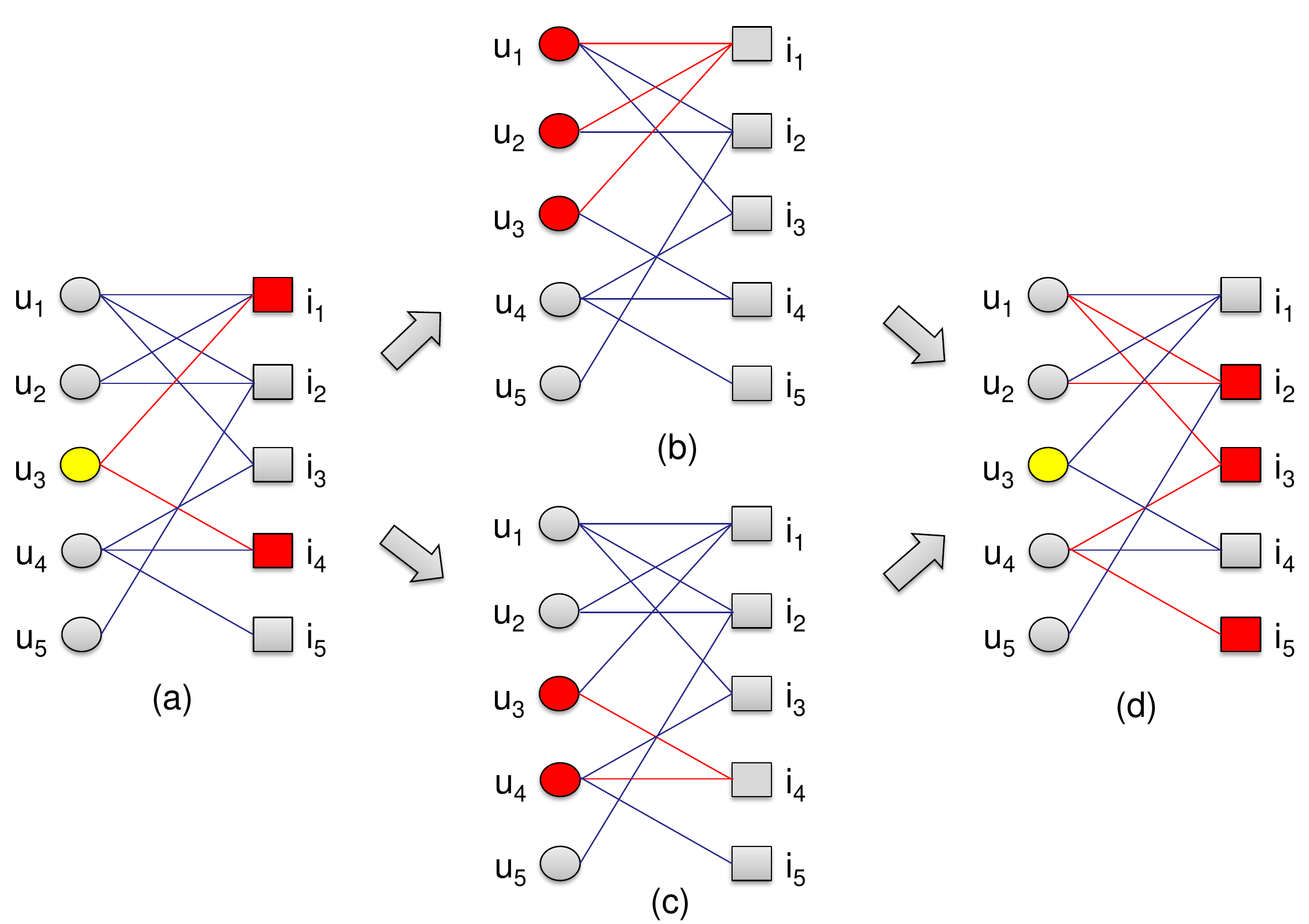}
	\caption{(Color online) The resource assignment process for the MD model. The user (circle) in yellow is the target user and rectangles denote items. The red rectangles in (a) denote items with initial resource. The red circles/rectangles in (b), (c) and (d) denote users/items who receive resource from neighbor items/users. 
		Following the resource transfer function, the final resource of candidate items ($i_1$, $i_2$ and $i_3$) are $0.28$, $0.28$ and $0.17$ in the MD model, and $0.30$, $0.31$ and $0.17$ for the MDEL model.}
	\label{fig:ProbS}
\end{figure}

Figure \ref{fig:ProbS} shows an example for the resource assignment process for the MD model on a user-item bipartite graph. $u_3$ is the target user that needs recommendations and the candidates for $u_3$ are his/her uncollected items $i_2$, $i_3$ and $i_5$. Figure~\ref{fig:ProbS}(a) shows the initial resource distribution, that the collected items $i_1$ and $i_4$ are assigned with resource of 1, and the other items are assigned with 0. In the first step (as shown in Figure~\ref{fig:ProbS}(b) and Figure~\ref{fig:ProbS}(c)), $i_1$ assigns its resource averagely  to its connected users $u_1$ (1/3), $u_2$ (1/3) and $u_3$ (1/3), and $i_4$ to $u_3$ (1/2) and $u_4$ (1/2). In the next step, $i_2$ would obtain resource from $u_1$ and $u_2$, $i_3$ from $u_1$ and $u_4$, $i_5$ from $u_4$, the amount of resource are all assigned from each node averagely to all its neighbors. Note that, although the collected items $i_1$ and $i_4$ of the target user $u_3$ are not painted red in Figure~\ref{fig:ProbS}(d), they would also get resource from the neighbor users, but as they are not candidates, we do not paint them in red color.

We then formulate this two-step resource redistribution process to an item-to-item transfer function, which writes as

\begin{center}
	\begin{equation}
	\label{equation-MD} 
	trans(i,j) = \frac{1}{k_{j}}\sum\nolimits_{v\in U}{\frac{a_{v i}a_{v j}}{k_{v}}}
	\end{equation}
\end{center}

\noindent where $j$ is the collected item of the target user, $i$ is one of the candidate items, and the resource is transferred from $j$ to $i$. The total resource that $i$ would get is the sum of resource from all the collected items.

\subsection{The ExTrA-based diffusion model}

In many applications, RSes involve side information, such as user profiles, item features, social trust information, or natural language comments, to help to predict which items the target user might prefer. For example, in news recommendation system, the latest news is usually more likely to be viewed than the earlier ones, then, in their RS models, they could set a time decay function to decrease the weights of news. However, incorporating extra information usually leads to the increase of computational cost, especially when involving complicated  data processing techniques, such as natural language processing, image processing and computer vision. In our model, to demonstrate the significant effect of highlighting fabricated experts in the methods based on bipartite network structures, we do not involve any side information, instead, we extensively mine users' tracks on items and extract those who have  diverse item preferences.  

To appropriately formulate one's capability in finding diverse items, we first qualitatively figure out what features these specified experts might have as the example shown in Table~\ref{tab:expert-feature}, corresponding to Figure~\ref{fig:ProbS}.

\begin{table}[!ht]\scriptsize
	\caption{Illustration of expertise level of users in Figure~\ref{fig:ProbS}. Active, Normal and Inactive are different levels of  activeness. Popular, Normal and Unpopular are the popularity of items.}
	\label{tab:expert-feature}
	
	\begin{tabular}{|p{0.05\columnwidth}|p{0.06\columnwidth}|p{0.07\columnwidth}|p{0.07\columnwidth}|p{0.08\columnwidth}|p{0.09\columnwidth}|p{0.09\columnwidth}|p{0.08\columnwidth}|}
		
		\hline
		\rowcolor{mygray}
		Users & Active & Normal & Inactive & Popular & Normal  & Unpopular & Expertise \\
		\rowcolor{mygray}
		&&&& ($i_1$,$i_2$) & ($i_3$, $i_4$) & ($i_5$)&  Level\\
		\hline
		$u_1$ & ${\surd}$ & & & ${\surd}$ & ${\surd}$&  & Higher\\
		\hline
		$u_2$ & & ${\surd}$& & ${\surd}$ & & & Normal\\
		\hline
		$u_3$ &  &${\surd}$ & & ${\surd}$ & ${\surd}$& & High\\
		\hline
		$u_4$ & ${\surd}$ & & &  &${\surd}$ & ${\surd}$ &Highest\\
		\hline
		$u_5$ &  &  & ${\surd}$ & ${\surd}$  & & & Low\\
		\hline
	\end{tabular}
\end{table}

At the very beginning, we believe that the active users who prefer selecting unpopular items (niches) have the best ability to improve the exposure of niches. To appropriately formulate it, Equation~\ref{equation-expert-function-better} is designed for extracting the experts who are active and could explore unpopular items (niches). If user $u$ have collected many unpopular items, his/her expertise level $e(u)$ will be the largest, otherwise, if $u$ have collected only a few popular items, $e(u)$ will be the smallest. Thus, $e(u)$ can evaluate the expertise level of users properly. Note that, we only improve the weights of expert users during the resource transfer process rather than pick these experts out from the bipartite graph~\cite{zeng2014uncovering}. Thus, we could improve the diversity,  without  a cliff-style decrease of the accuracy which happens strikingly in model HC.

\begin{center}
	\begin{equation}
	\label{equation-expert-function-better} 
	e(u) = \displaystyle{\sum\nolimits_{i\in I}}{\frac{a_{u i}}{k_i}}
	\end{equation}
\end{center}

Next we show how to incorporate the expertise level of users $e(u)$ into the resource transfer process. That is, in the first step, not assigning the resource of an item averagely to all its neighbor users, but proportional to the expertise values of neighbor users. For example, in Figure~\ref{fig:ProbS}(b), the resource of $u_1$ received from $i_1$ is $\frac{e(u_1)}{e(u_1) + e(u_2) + e(u_3)}$, similar for $u_2$ and $u_3$. The final transfer function could be written as:

\begin{center}
	\begin{equation}
	\label{equation-expert-final} 
	trans(i,j) = \frac{1}{k_{j}} {\sum\nolimits_{u\in U}}{a_{u i} a_{u j} N(e(u))},
	\end{equation}
\end{center}
where 
\begin{center}
	\begin{equation}
	\label{equation-expert-module} 
	N(e(u)) =  (\frac{e(u)}{\sum_{v\in U}{a_{vj}e(v)}})^\lambda
	\end{equation}
\end{center}
represents the resource percentage that user $u$ will get from item $j$, with an adjustable exponential parameter $\lambda$.

We found that the $e(u)$ formula defined above does not work well for improving diversity. For the real-world data, most users would be inactive users, thus, this function has effect on only a small part of users, in other words, can distinguish only a few users with others. In this way, the overall performance improvement would not happen. 
Instead, 
we find that taking into consideration the average degree of selected items for a user will recruit more users (see Figure~\ref{fig:user_degree_vs_large_eng:a}). For example, if user $u$ and $v$ are both of degree 2, and user $u$ has selected one niche item and one popular item, but user $v$ has selected two niches. In the case of above $e(u)$, $u$ and $v$ would obtain almost the same  expertise value. However,  $u$ is clearly more capable of finding diverse items. Thus, we design another applicable $e(u)$ formula, which is defined as the average value of the popularity of collected items of users.  This \underline{M}ass \underline{D}iffusion model with \underline{E}xperts collecting \underline{L}arge-degree items is short as {\bf MDEL}. This kind of expert neighbors transfer resource to a larger range of items, from popular to unpopular ones (see Equation~\ref{equation-expert-function-best}). 

\vspace{-15pt}
\begin{center}
	\begin{equation}
	\label{equation-expert-function-best} 
	e(u) = \frac{\displaystyle{\sum\nolimits_{i\in I}}{a_{u i}{k_i}}}{\displaystyle{\sum\nolimits_{i\in I}}{a_{u i}}}
	\end{equation}
\end{center}

\subsection{Data sets for performance evaluation}

Hear we introduce three data sets used in this paper. MovieLens is the data set used in this subsection, which was collected by the GroupLens Research Project at the University of Minnesota and can be found at the website\footnote{https://grouplens.org/datasets/movielens/}. The other two real-world datasets we will use in the later sections are Netflix and RYM. Netflix~\cite{bennett2007netflix} is a randomly sampled subset of the huge data set provided by the Netflix company for the Netflix Prize\footnote{www.netflixprize.com}. RYM is obtained by downloading publicly available data from the music ratings website\footnote{www.RateYourMusic.com}. In this paper, we make use of nothing but the binary information whether there exists an
interaction or explicit  preference between a user and an item in the past. 
The datasets and experiment codes are released to facilitate the research community\footnote{https://github.com/anyahui120/ExTRA-Expert-track-based-Recommendation-Algorithm}.

In our experiments, each data set is randomly divided into two subsets: the training set $E^T$, and the probe set $E^P$. We name the dataset with the title and the percentage of training set. For example, on MovieLens, if the size of $E^T$ is $80\%$ and $E^P$ is $20\%$, we represent it as MovieLens($E^T$=80). Training set is treated as known information, which is also used for extracting the specific experts, and the testing set is used to  evaluate the performance of different methods. The statistics of three datasets are presented in Table~\ref{tab:datasets}.

\begin{table}[!ht] 
	\centering
	\caption{The basic statistics of three data sets, where $\#users$, $\# items$ and $\#links$ denote the number of users, items and edges, respectively; $\langle{k_u}\rangle$ and $\langle{k_i}\rangle$ are the average degrees of users and items. Degree of a user is the number of items the user collected, which is also defined as {\bf activeness}. Degree of an item is the number of users that have collected the target item, which is defined as {\bf popularity}.}
	\label{tab:datasets}
	\begin{tabular}{|c|c|c|c|c|c|}
		\hline
		\rowcolor{mygray}
		Data set&$\#users$ & $\#items$ & $\#links$ & $\langle{k_u}\rangle$ & $\langle{k_i}\rangle$ \\
		\hline
		MovieLens & 943 & 1,682 & 100,000 & 106 & 59.5 \\
		\hline
		Netflix & 10,000 & 5,640 & 701,947 & 70.2 & 124.5 \\
		\hline
		RYM & 33,762 & 5,267 & 675,817 & 20 & 128.3 \\
		\hline
	\end{tabular} 
\end{table}

\subsection{Primary effect of an ExTrA-based model}
\label{sub:trade-off}
In this subsection, a simple experiment is conducted on the MovieLens data set to show the diversity-accuracy performance of the MDEL model, compared with the original MD model.

\begin{figure*}[!htbp]
	\centering
	\subfigure[The dense data set]{
		\includegraphics[width=0.9\columnwidth]{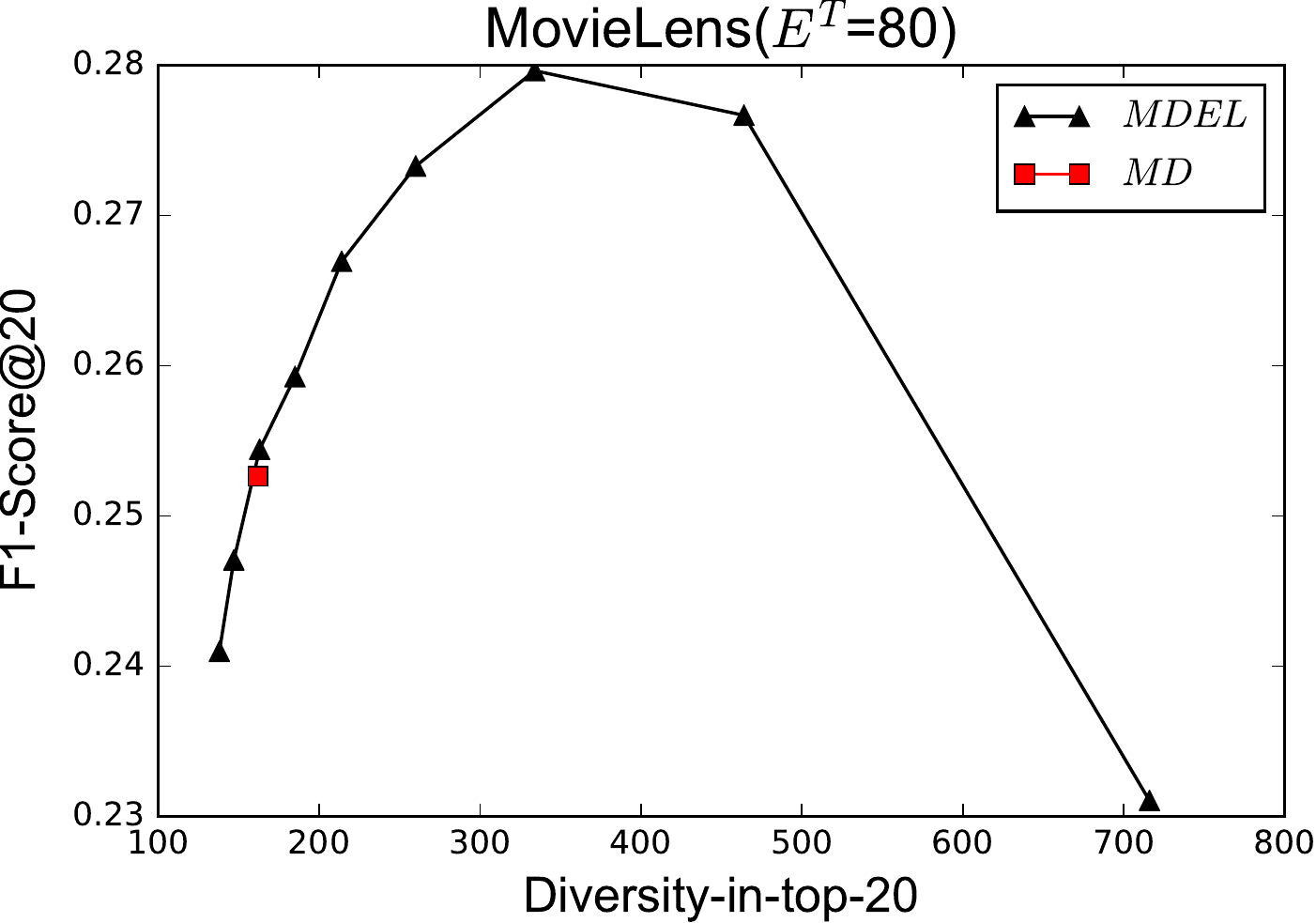}}
	\subfigure[The sparse data set]{
		\includegraphics[width=0.9\columnwidth]{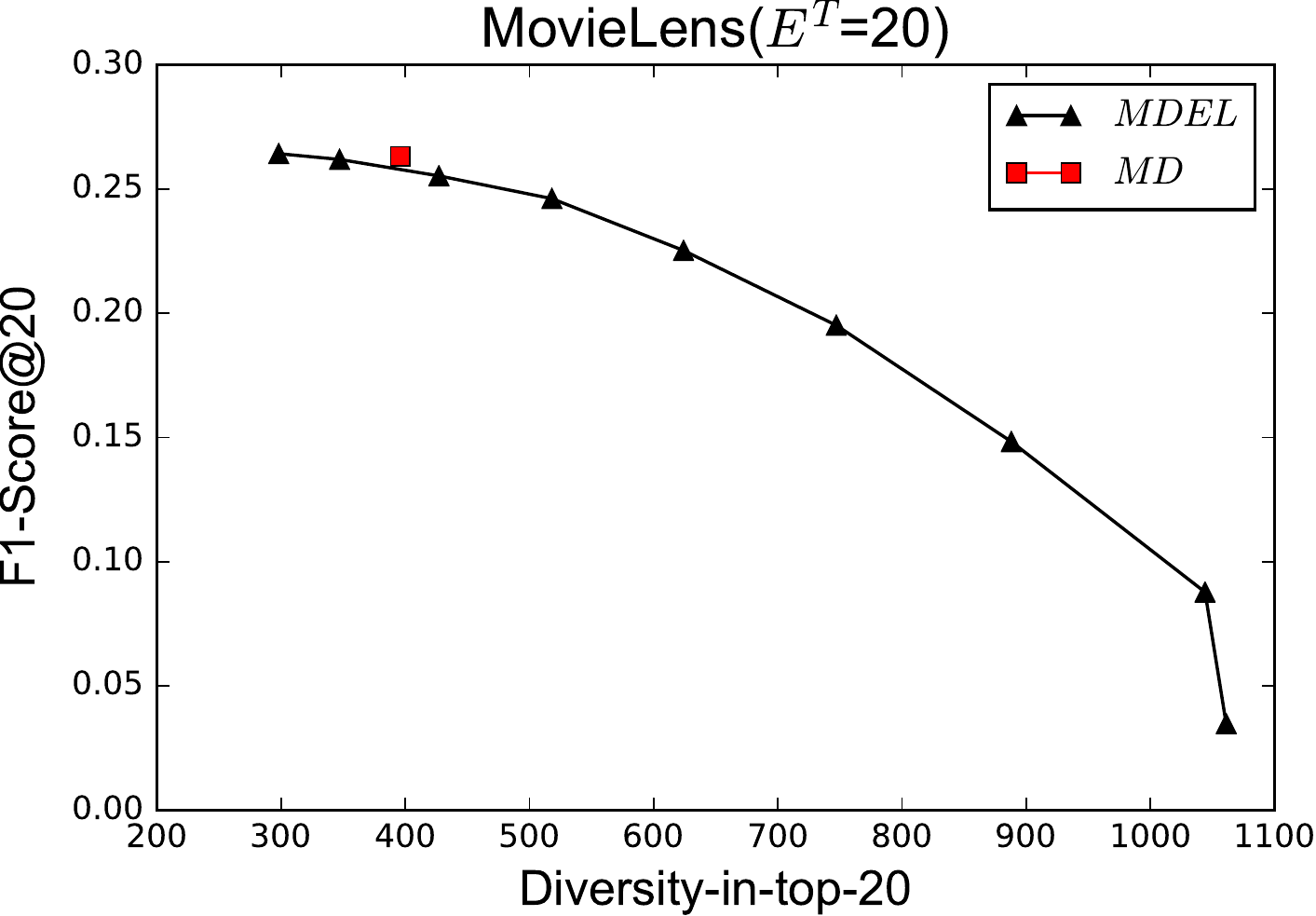}}
	\caption{(Color online) The accuracy vs diversity performance of standard diffusion-based method MD (red square) and an ExTrA-based MD method MDEL (black triangle) on MovieLens($E^T$=80) and MovieLens($E^T$=20), where the recommendation length is 20. The points on each line are the values for a range of $\lambda$, from 0.1 to 0.9.}
	\label{fig:accuracy-diversity-traseoff}
\end{figure*}

We compare the performance of MDEL and the standard MD models on MovieLens dataset, andthis comparison result  is presented  as the accuracy-diversity plot in Figure~\ref{fig:accuracy-diversity-traseoff}. The accuracy and diversity metrics we used here is described in Section \ref{subsec:metrics}. Particularly, the first panel of Figure~\ref{fig:accuracy-diversity-traseoff} (MovieLens($E^T$=80)) shows that, compared to the standard MD method (red square), MDEL increases recommendation diversity from 162 to 716 ($\lambda$=0.9); however, the recommendation accuracy is dropped from 25.3\% to 23.1\%. In this case, diversity is gained 342\% with a little loss of accuracy (8.7\%), however, with a proper $\lambda$ (from 0.3 to 0.8), both the accuracy and diversity would be improved (when $\lambda=0.7$, $x = 464$ and $y = 27.7\%$). 
But in the other cold start dataset, MovieLens($E^T$=20), shown in the second panel of Figure~\ref{fig:accuracy-diversity-traseoff}, despite the significant diversity gain from 396 to 1061 (+167.9\%), such a significant accuracy loss (from 26.3\% to 3.4\%) would not be acceptable in most real-life personalized applications. Therefore, in real applications, the trade-off between accuracy and diversity by adjusting parameter $\lambda$ allows to achieve significant diversity gains while bounding accuracy loss, which depends on how much accuracy loss is tolerable in a given application.

\section{Several other Expert extraction methods}
\label{sec:additional}

The above simple experiment on MovieLens shows that MDEL is effective for our objective. Next We will introduce several other Expert extraction methods inspired by different motivations, from simple to complicated ones, and check the distribution of expertise level for each method on three data sets.
Finally, we will check their abilities of 
improving recommendation diversity when combined with the MD model.

\begin{figure*}[!htbp]
	\centering
	\subfigure[The MDActivity model]{
		\label{fig:user_degree_vs_activity:a}
		\includegraphics[width=0.6\columnwidth]{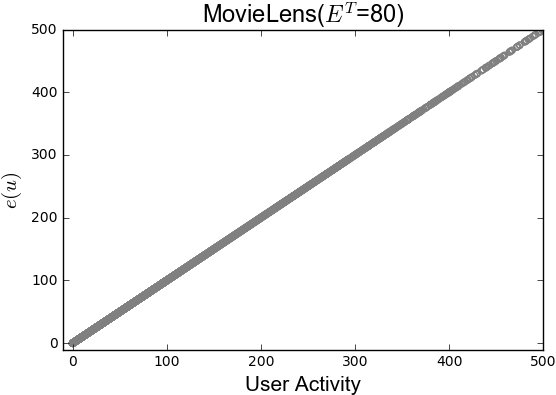}
		\includegraphics[width=0.6\columnwidth]{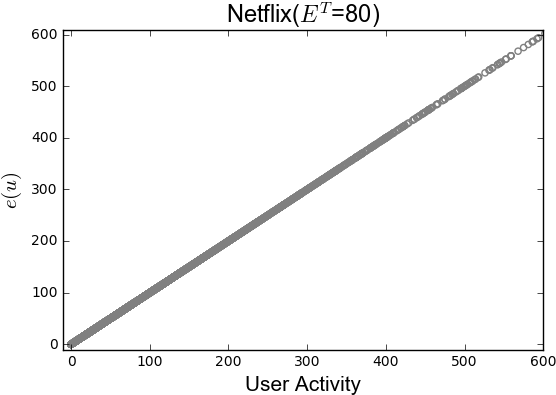}
		\includegraphics[width=0.6\columnwidth]{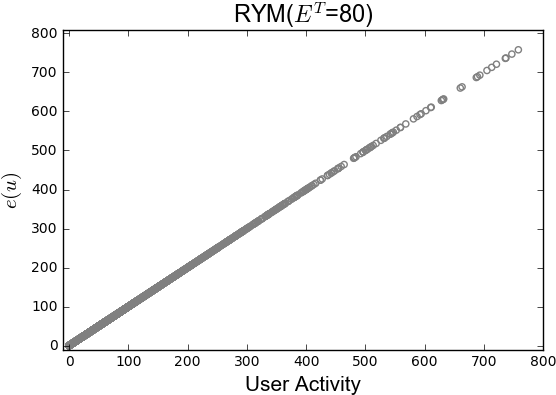}}
	
	\subfigure[The MDEL model]{
		\label{fig:user_degree_vs_large_eng:a}
		\includegraphics[width=0.6\columnwidth]{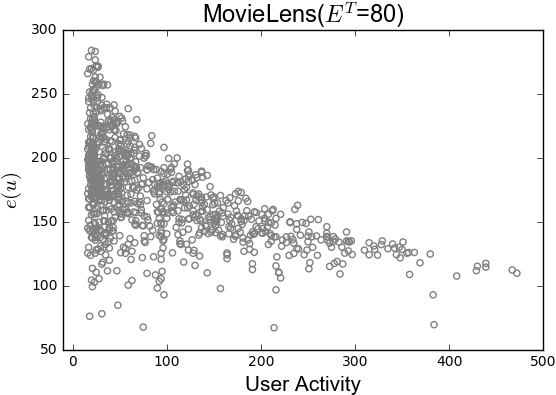}
		\includegraphics[width=0.6\columnwidth]{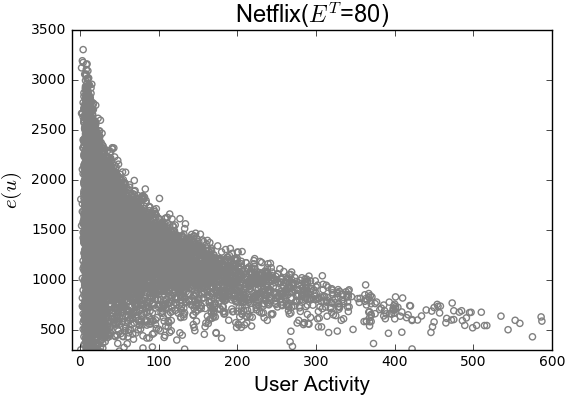}
		\includegraphics[width=0.6\columnwidth]{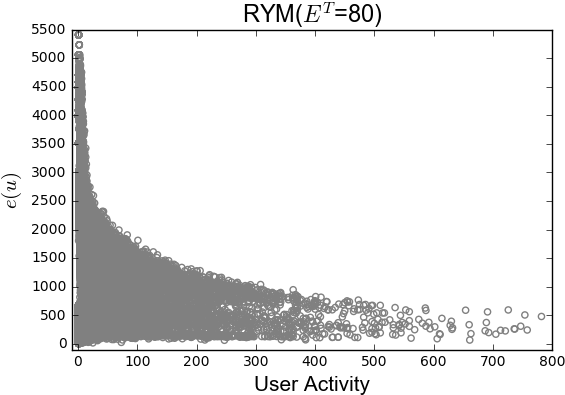}}

	\subfigure[The MDGini model]{
		\label{fig:user_degree_vs_gini:a}
		\includegraphics[width=0.6\columnwidth]{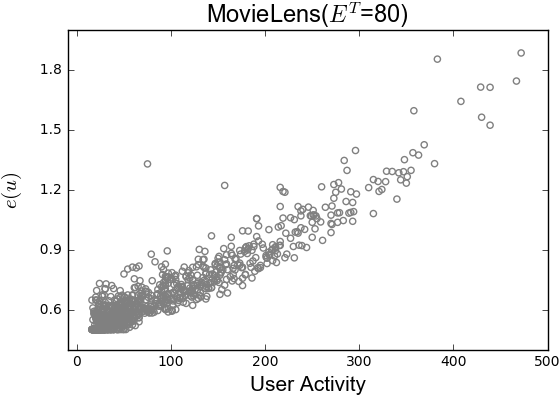}
		\includegraphics[width=0.6\columnwidth]{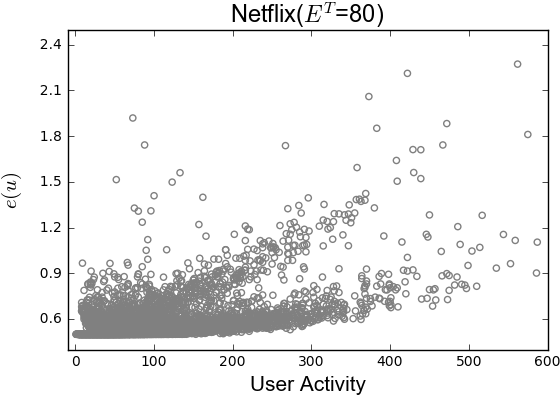}
		\includegraphics[width=0.6\columnwidth]{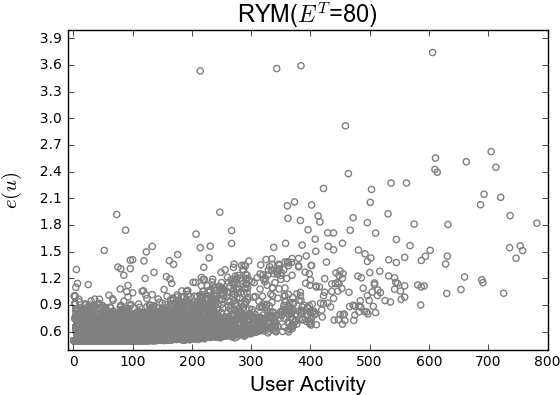}}
	
	\subfigure[The MDSim  model]{
		\label{fig:user_degree_vs_sim:a}
		\includegraphics[width=0.6\columnwidth]{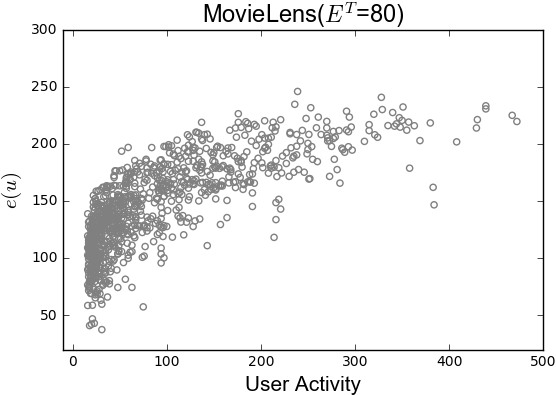}
		\includegraphics[width=0.6\columnwidth]{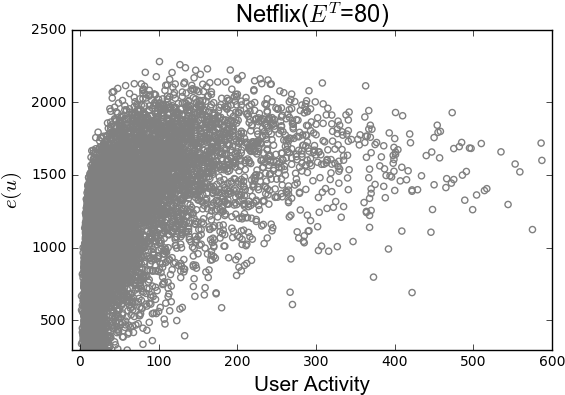}
		\includegraphics[width=0.6\columnwidth]{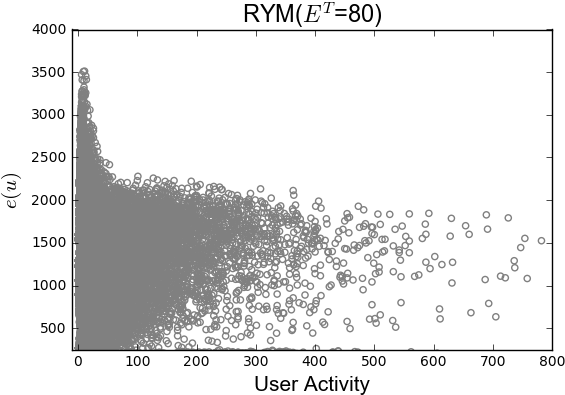}}
	
	\subfigure[The  MDSim2 model]{
		\label{fig:user_degree_vs_sim2:a}
		\includegraphics[width=0.6\columnwidth]{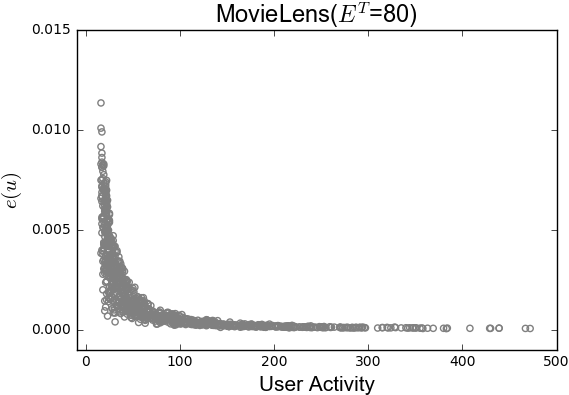}
		\includegraphics[width=0.6\columnwidth]{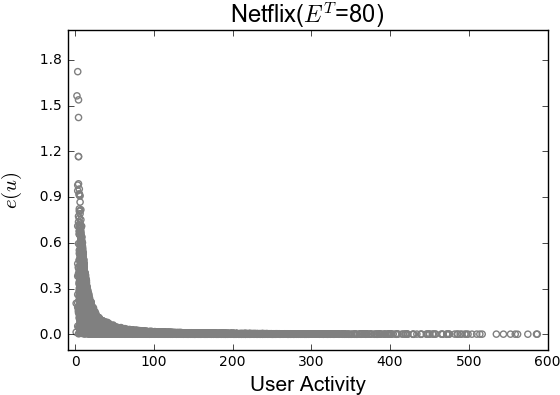}
		\includegraphics[width=0.6\columnwidth]{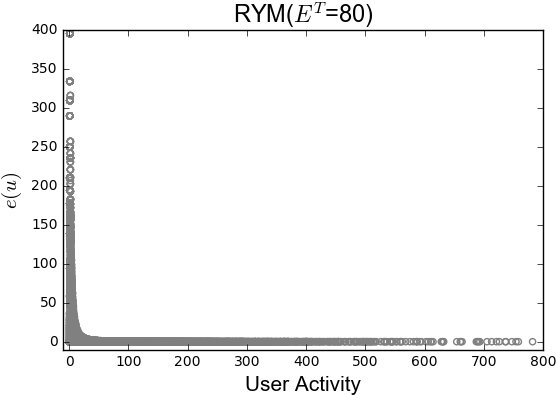}}
	\caption{The expertise distribution for 5 ExTrA-based models. The $x$ axis represents the activity of users and the $y$ axis represents the values of user expertise. For each subfigure, the datasets from left to right are MovieLens($E^T$=80), Netflix($E^T$=80) and RYM($E^T$=80), respectively.}
	\label{fig:user_energy_dis}
\end{figure*}

First, we employ the most simple idea, that is to extract the most \underline{active} users as experts and apply it in the MD model, which is called {\bf MDActivity}, and regard it as the baseline of other well-designed ExTrA-based models. Therefore, we simply use the activity level as the expertise level of users. MDActivity is formulated as:

\begin{itemize}
	\item {\bf MDActivity: Active in History Data}, i.e., the expertise for a user is the activity level of the user. More formally:
	$e(u) = \displaystyle{\sum\nolimits_{j\in I}{a_{u j}}},$
\end{itemize}
The distribution of $e(u)$ values of different users is shown in Figure~\ref{fig:user_degree_vs_activity:a}, which is a diagonal line.

Next we propose 3 delicately designed methods of extracting the experts, considering the Gini coefficient  of popularity  of collected item  by the user ({\bf MDGini}), the similarity of the user to all other users ({\bf MDSim}), the similarity of the user to all other users divided by his/her activity level ({\bf MDSim2}).

In economics, the Gini coefficient, sometimes called Gini index, or Gini ratio, is a measure of inequality of the income or wealth distribution of a nation's residents. In MDGini, we use Gini coefficient of popularity of collected items as the expertise value of users in the process of diffusion, where higher Gini coefficient means more diverse item-popularity preference of one user:

\begin{itemize}
	\item {\bf MDGini: Diversity of Preference}, i.e, gini-coefficient of popularity of user' selected items:
	
	$e(u) = 2\sum\nolimits_{k=1}^{|I_u|}{[(\frac{|I_u|+1-i_k}{|I_u|+1})*(\frac{count(i_k)}{|I_u|})]}$, where $I_u$ is the item set that $u$ have selected, $k$ is the rank of item $i_k$, sorted by the popularity of items in an ascending order, count($i_k$) is the count of items that have the same rank in the rank list (which means that they have the same popularity).  
\end{itemize}
The distribution of $e(u)$ values of different users for MDGini is shown in Figure~3(c).

MDEL, MDActivity and MDGini are three typical methods that extract features based on only each user's collection records on popular/unpopular items. We already know that the classic item-based Collaborative Filtering (CF) method is based on the similarity between items, which treats users indistinguishably. Similarly, the MD model, in the first step of the resource diffusion, also assigns users connected with the same item with equal amount of resource. However, if we distribute the resource to users based on their similarity to the target user, neighbors who are more similar would obtain more resource. We call this ExTrA-based model as MDSim. The biggest difference of MDSim from MDActivity is that MDSim would recruit more users whose activity are in middle level, as shown in Figure~\ref{fig:user_degree_vs_sim:a}. 

\begin{itemize}
	\item {\bf MDSim: More Similar to Others}, i.e., the expertise value of a user $u$ is the sum of cosine similarity to all the other users:
	$e(u) = \sum\limits_{v \in {\bf U}, v \ne u} \frac{|I_v \bigcap I_{u}|}{\sqrt{|I_v||I_{u}|}}$, where $I_{u}$ and $I_v$ are the item sets that user $u$ and $v$ have selected.
\end{itemize}

Based on MDSim, we further penalize the similarity between the active user pair to increase the importance of the similar but not that active users:

\begin{itemize}
	\item {\bf MDSim2: More Similar to Inactive users}, i.e., the similarity between the active users are penalized:
	
	$e(u) = \sum\limits_{v \in {\bf U}, v \ne u} \frac{|I_v \bigcap I_{u}|}{(|I_v||I_{u}|)^2}$,
\end{itemize}

Figure~\ref{fig:user_energy_dis} presents the distribution of expertise level for each method on three data sets, with  sharply different shapes. Cleartly, the distribution of expertise level against user activity for MDActivity is a diagonal line (Figure~\ref{fig:user_degree_vs_activity:a}), which is used for comparing baseline with other models. From the macro perspective, the distribution of expertise are of different shapes in Figure~\ref{fig:user_energy_dis}(a-e). Figure~\ref{fig:user_degree_vs_activity:a}, \ref{fig:user_degree_vs_gini:a} and \ref{fig:user_degree_vs_sim:a} show positive correlations between the expertise and user activity, while Figure~\ref{fig:user_degree_vs_large_eng:a} and \ref{fig:user_degree_vs_sim2:a} with negative correlations. This is because the effects of user activity are not that significant in MDEL and MDSim2. In the MD models, the user activity plays an important role in the resource transfer function. Thus, even with the similar distribution shape, the slope of expertise values for MDSim2 is sharper than MDEL, which would further weaken the effect of user activity and increase the effect of users with diverse selections, no matter they are active or not.

Next we tested these 5 ExTrA-based MD models on three datasets to see the effectiveness and robustness in improving the diversity, using the standard MD model as the baseline. the test is conducted on a fix sparsity ($E^T$=80) for each dataset. 
The performance of each proposed ExTrA-based method is measured in terms of F1-Score@K and Diversity-in-top-K (see Sec.~\ref{subsec:metrics}), and, since there is no criterion for the trade-off between accuracy and diversity, we empirically select some typical values of $\lambda$ (in order to save the space) to show the increasing/decreasing tendency for both accuracy and diversity in Table~\ref{tab:final results} for MovieLens($E^T$=80), Netflix($E^T$=80) and RYM($E^T$=80).

\noindent
\begin{table*}[!ht]
	\centering
	\caption{The accuracy vs diversity performance of 5 ExTrA-based methods on MovieLens($E^T$=80), Netflix($E^T$=80) and RYM($E^T$=80). The value of the parameter $\lambda$ is listed in the first column.
		The two values in each entry are F1-Score@20 and Diversity-in-top-20. In particular, MD: (0.253, 162), MD: (0.184,262) and MD: (0.169,2704) are for the original MD model as the baseline for each dataset.}
	\label{tab:final results}
	\begin{tabular}{|c|c|c c|c c|c c|c c|c c|}
		\hline
		\rowcolor{mygray}
		Datasets &$\lambda$ & \multicolumn{2}{|c|}{MDEL} & \multicolumn{2}{|c|}{MDActivity} & \multicolumn{2}{|c|}{MDGini} & \multicolumn{2}{|c|}{MDSim} & \multicolumn{2}{|c|}{MDSim2} \\
		\hline
		\multirow{5}{*}{MovieLens($E^T$=80)}
		&0.5 & 0.267 & 214  & 0.260 & 183 & 0.263 & 195 & 0.264 & 196 & 0.279 & 286 \\ \cline{2-12}
		& 0.6 & 0.273 & 260  & 0.266 & 215 & 0.269 & 230 & 0.269 & 235 & 0.286 & 345 \\ \cline{2-12}
		&0.7 & 0.280 & 334  & 0.270 & 264 & 0.274 & 292 & 0.275 & 295 & \textbf{0.287} & \textbf{426}\\ \cline{2-12}
		\textbf{MD: (0.253, 162)}&0.8 & 0.277 & 464  & 0.247 & 357 & 0.272 & 399 & 0.271 & 414 & 0.281 & 650\\ \cline{2-12}
		&0.9 & 0.231 &  716 & 0.202 & 550 & 0.227 & 595 & 0.224 & 619 & \textbf{0.227} & \textbf{911} \\ \hline\hline

		\multirow{5}{*}{Netflix($E^T$=80)}
		&0.5 & 0.191 & 493  & 0.186 & 363 & 0.189 & 446 & 0.187 & 420 & 0.200 & 724 \\ \cline{2-12}
		&0.6 & 0.196 & 744  & 0.191 & 538 & 0.194 & 655 & 0.192 & 643 & 0.204 & 1120\\
		\cline{2-12}
		&0.7 & 0.202 & 1100 &  0.196 & 800 & 0.200 & 978 & 0.197 & 927 & \textbf{0.209} & \textbf{1625} \\
		\cline{2-12}
		\textbf{MD: (0.184, 262)}&0.8 & 0.204 & 1503 & 0.193 & 1052 &0.202  & 1314 & 0.200 & 1285 & 0.204 & 2101 \\
		\cline{2-12}
		&0.9 & 0.155 & 1696 & 0.077 & 1111 & 0.145 & 1468 & 0.145 & 1433 & \textbf{0.135} & \textbf{2320}\\
		\hline \hline
		\multirow{5}{*}{RYM($E^T$=80)}
		&0.5 & 0.154 & 4288  & 0.147 & 3524 & 0.155 & 3924 & 0.149 & 4037 & 0.150 & 4641 \\
		\cline{2-12}
		&0.6 & 0.157 & 4574& 0.148 & 3835 & 0.158 & 4205 & 0.152 & 4373 & 0.145 & 4809\\
		\cline{2-12}
		&0.7 & 0.158 & 4780  & 0.150 & 3994 & \textbf{0.161} & \textbf{4417} & 0.154 & 4575 & 0.133 & 4932 \\
		\cline{2-12}
		\textbf{MD: (0.169, 2704)}&0.8 & 0.156 & 4841  & 0.147 & 4004 & \textbf{0.161} & \textbf{4518} & 0.154 & 4653 & 0.113 & 4994 \\
		\cline{2-12}
		&0.9 & 0.137 & 4807 & 0.123 & 3791 & 0.153 & 4491 & 0.143 & 4556 & \textbf{0.066} & \textbf{5000} \\
		\hline
	\end{tabular}
\end{table*}

The overall performance is consistent with the trade-off of accuracy and diversity discussed in subsection~\ref{sub:trade-off}. As $\lambda$ changes from 0.5 to 0.9, the accuracy for each proposed methods follows the shape of first increasing and then decreasing, one except case is the MDSim2 on the RYM data set, which also follows the same shape if we show the result from $\lambda=0.1$ to $\lambda =0.9$. While, for MovieLens and Netflix, the diversity of each method keeps increasing, which is significantly improved, and for RYM, the improvement of diversity for each method is not significant, but still has a increasing shape. Therefore, choosing a proper $\lambda$  allows the system to improve both accuracy and diversity in some cases, or at least achieve a desired balance between accuracy and diversity. 

In particular, if we compare the 5 ExTrA-based methods with the original MD model in the potential of diversity improvement and the ability to keep accuracy when improving diversity, on all three datasets, they show the similar performance ranking sequence : MDSim2 \textgreater MDEL \textgreater MDSim \textgreater MDGini \textgreater MDActivity  \textgreater MD. MDSim2 gets the best overall performance, which achieves the best accuracy and diversity compared to all the other methods, MDEL is ranked the second best, MDSim and MDGini are similar to each other and ranked the third and fourth positions, and, MDActivity is the worst one but still much better than the original MD model.

\section{Performance comparison with existing models}
\label{sec:analysis}
\subsection{Two typical diffusion variant models}

We have evaluated the performance of the proposed 5 ExTrA-based methods with comparisons with the original MD model. To further validate the diversity performance of the ExTrA-based methods extensively, we will compare them with two state-of-the-art MD-improved models, HHP and BHC, which were proposed to improve simutaneously the diversity and accuracy and also work based on the history behavior data without involving any side-information. Note that, we could also apply the best expert extraction method to HHP and BHC, which could be named as HHPSim2 and BHCSim2, however, that would bring more computing cost because of introducing more parameters. Therefore, we only compare the ExTrA-based methods with standard HHP and BHC models to show the effectiveness of fabricated experts.

\begin{itemize}
	\item HHP is a nonlinear hybrid of MD and HC models, which tries to solve the dilemma of accuracy and diversity and increases both the accuracy and diversity.  
	\item BHC is a biased Heat Conduction model, which tries to reduce the bias that niche items absorb more resource than the popular ones, which leads to very poor accuracy in HC. This method  compensates the resource absorbed by popular items in the last step of the resource propagation.
\end{itemize}

The performance of HHP and BHC on solving the dilemma of accuracy and diversity are significant, however, the performance would degrade when the dataset is very sparse. The performance of HHP and BHC, comparing with MD are shown in Table~\ref{tab:netflix}. We can see that, compared with MD, the diversity has been improved significantly for Netflix($E^T$=80). However, for the cold start dataset ($E^T$=20), although HHP and BHC still work, the improvement for accuracy (F1-Score@20) and diversity (Diversity@20), on Netflix($E^T$=20), degrades to some extent. For example, on Netflix($E^T$=80), the improvement percentage of Diversity@20 for HHP compared to MD is 590\%, however $78\%$ on Netflix($E^T$=20). Note that, the  values of Precision@20 on Netflix($E^T$=20) are larger than those on Netflix($E^T$=80), because of the size of probe data (for the same dataset, usually the larger is the size of probe set, the larger is the precision).

\noindent
\begin{table*}[!ht]
	\centering
	\caption{The accuracy and diversity performance of the MD, HHP and BHC models on Netflix($E^T$=80) and Netflix($E^T$=20).}
	\label{tab:netflix}
	\begin{tabular}{|c|c|c|c|c|c|c|}
		\hline
		\rowcolor{mygray}
		\hline
		Data & Methods & $\lambda_{opt}$ & Precision@20  & Recall@20 &F1-Score@20 & Diversity@20 \\
		\hline
		\multirow{3}{*}{Netflix($E^T$=80)} 
		&MD&NA&0.140&0.269&0.184&262 \\  \cline{2-7}
		&HHP&0.8&0.161&0.299&0.209&1809 \\ \cline{2-7}
		&BHC&0.8&0.156&0.293&0.203&1454 \\ \hline
		\hline
		\multirow{3}{*}{Netflix($E^T$=20)}&MD&NA&0.360&0.197&0.254&1708 \\ \cline{2-7}
		&HHP&0.3&0.369&0.202&0.261&3038 \\ \cline{2-7}
		&BHC&0.3&0.362&0.202&0.259&2244 \\ \hline
	\end{tabular}
\end{table*}

\subsection{Metrics of performance evaluation }
\label{subsec:metrics}

In the above, we measure the recommendation diversity by the total number of distinct items that are recommended across all users. It is necessary to measure whether each user gets a more diverse recommendation list. Thus, in this part, we also use two more metrics to measure the intra- and inter-list diversity. All the metrics that are used in this paper are list below:

\begin{itemize}
	\item {\bf Accuracy:} We assess the relevance of ranked items with {\bf Precision@K}, {\bf Recall@K} and {\bf F1-Score@K}.  Precision@K counts the number of hits among the top-K items of the recommendation list. Recall@K is the fraction of items (user likes) that have been retrieved over the total amount of relevant items. For real application scenario, users typically only see a few recommendations, thus, we set K as 10, 20, 30, 40, 50.
	
	\item {\bf Diversity-in-top-K:} The diversity metric we use in Sec.~\ref{sub:trade-off} is called \textbf{Diversity-in-top-K}, which is defined as the total number of distinct items in the recommendation lists of all users of length $K$~\cite{adomavicius2011improving}, also known as the Coverage diversity. 
	
	\item {\bf Intra-Diversity:} We also measure the recommendation diversity for a single user by intra-diversity (\textbf{intraD-I@K} for short) , which is based on a concept of intra-similarity defined as~\cite{lu2011information}:
	
	\begin{center}
		\begin{equation}
		\label{equ_ii} 
		I_{u} (K) = \frac{1}{K(K-1)} \sum\limits_{i\not= j, i,j\in I_u(K)} s_{ij},
		\end{equation}
	\end{center}
	where $ s_{ij}$ is the similarity between items $i$ and $j$, which in our case is represented by the cosine similarity. The average value of $I_{u} (K)$ on all users is the system's intra-similarity.  A good recommendation algorithm is expected to give fruitful recommendations and has the ability to guide or help the users to exploit their potential interest fields, and thus leads to a lower intra-similarity, i.e., higher intra-diversity. We assign the value of one minus $intra-similarity$ to the final value of intraD-I@K.

	\item {\bf Inter-Diversity:}Besides intra-diversity, the inter-diversity (also known as aggregative diversity), which considers the difference between the recommendation lists of each user pair, should be taken into consideration. Here we use hamming distance (\textbf{HD@K} for short) to evaluate it. Borrowing inspiration from the hamming distance between two strings~\cite{Zhou_2008}, the diversity is calculated in a similar way:
	\begin{center}
		\begin{equation}
		\label{equ_ij} 
		HD_{u,v} (K) = 1 -\frac{q_{uv}}{K},
		\end{equation}
	\end{center}
	
	where $q_{uv}$ is the number of common items in the top $K$ positions of both lists of user $u$ and user $v$. Clearly, if user $u$ and $v$ receive the same recommendation list, $HD_{uv}(K)$ = 0, while if their lists are completely different, $HD_{uv}(K)$ = 1. Averaging $HD_{uv}(K)$ over all pairs of active users in the probe set, we obtain the hamming distance $HD@K$ of the whole system, where greater value means better personalization of users' recommendation lists.
	
\end{itemize}

\subsection{Diversity comparison with existing models}

\begin{figure*}[ht]
	
	\centering
	
	\subfigure[The intra-diversity]{ 
		\label{fig:subfig:b} 
		\includegraphics[width=0.32\textwidth]{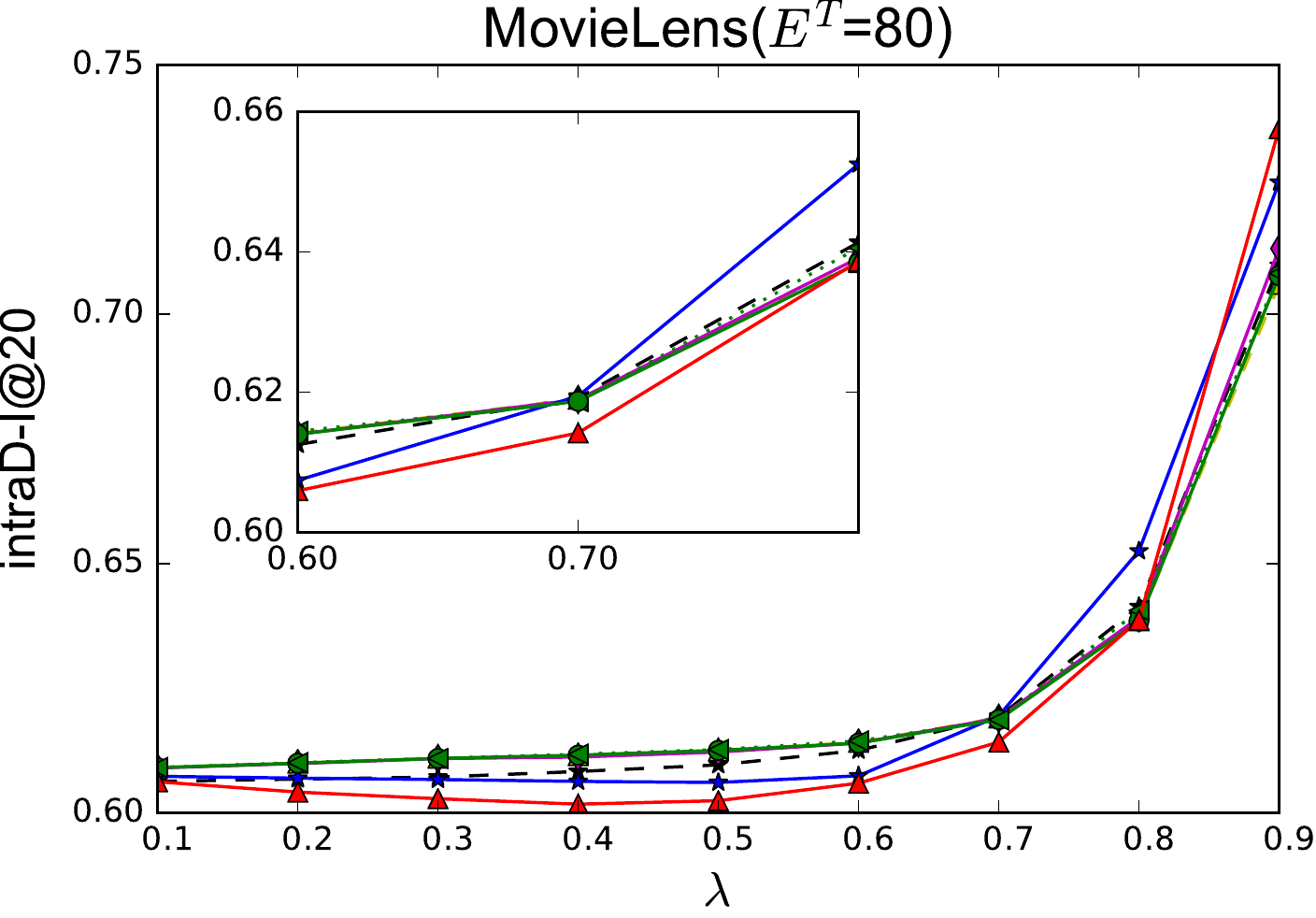}
		\includegraphics[width=0.32\textwidth]{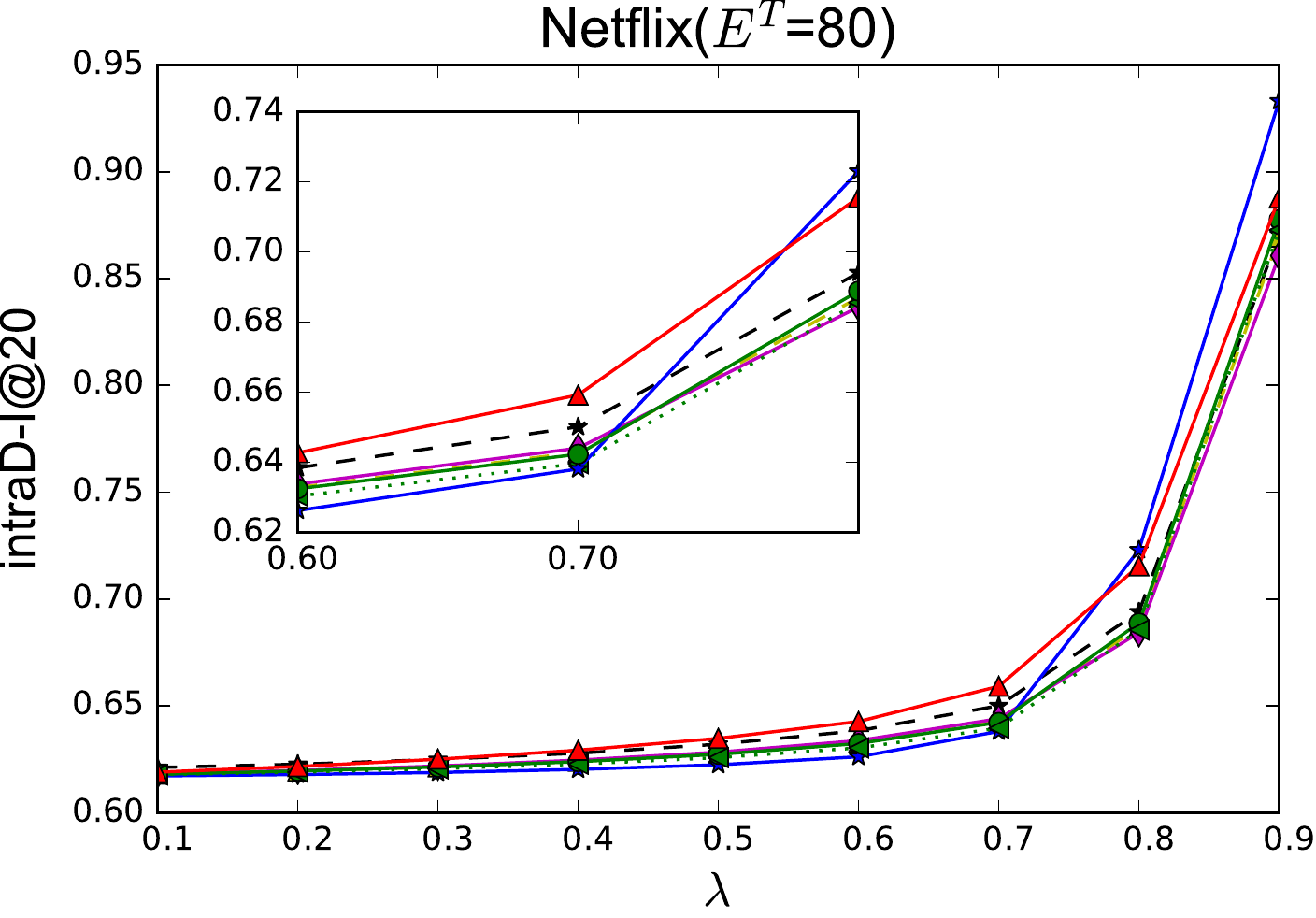}
		\includegraphics[width=0.32\textwidth]{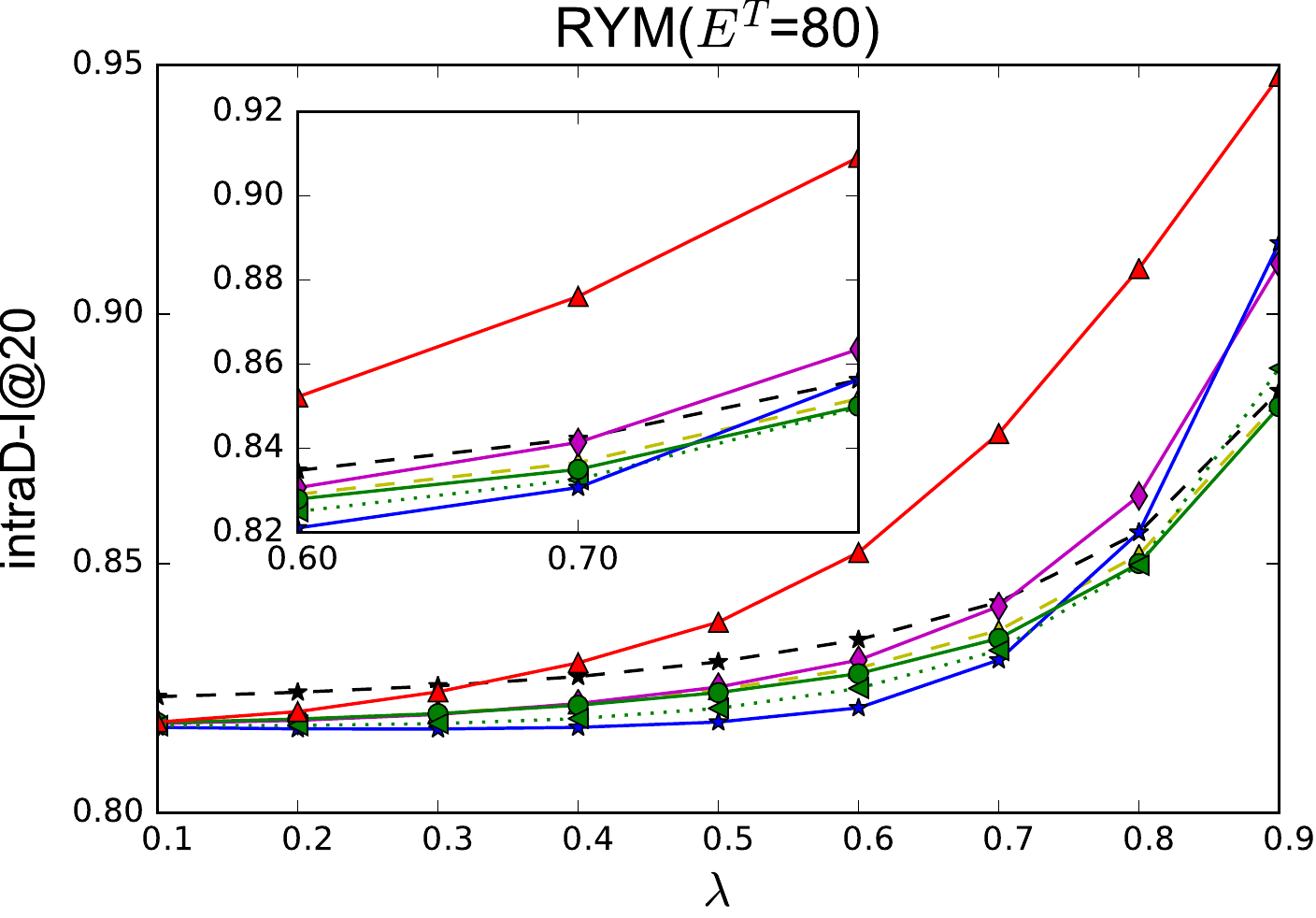}}
	
	\subfigure[The inter-diversity]{ 
		\label{fig:subfig:a} 
		\includegraphics[width=0.32\textwidth]{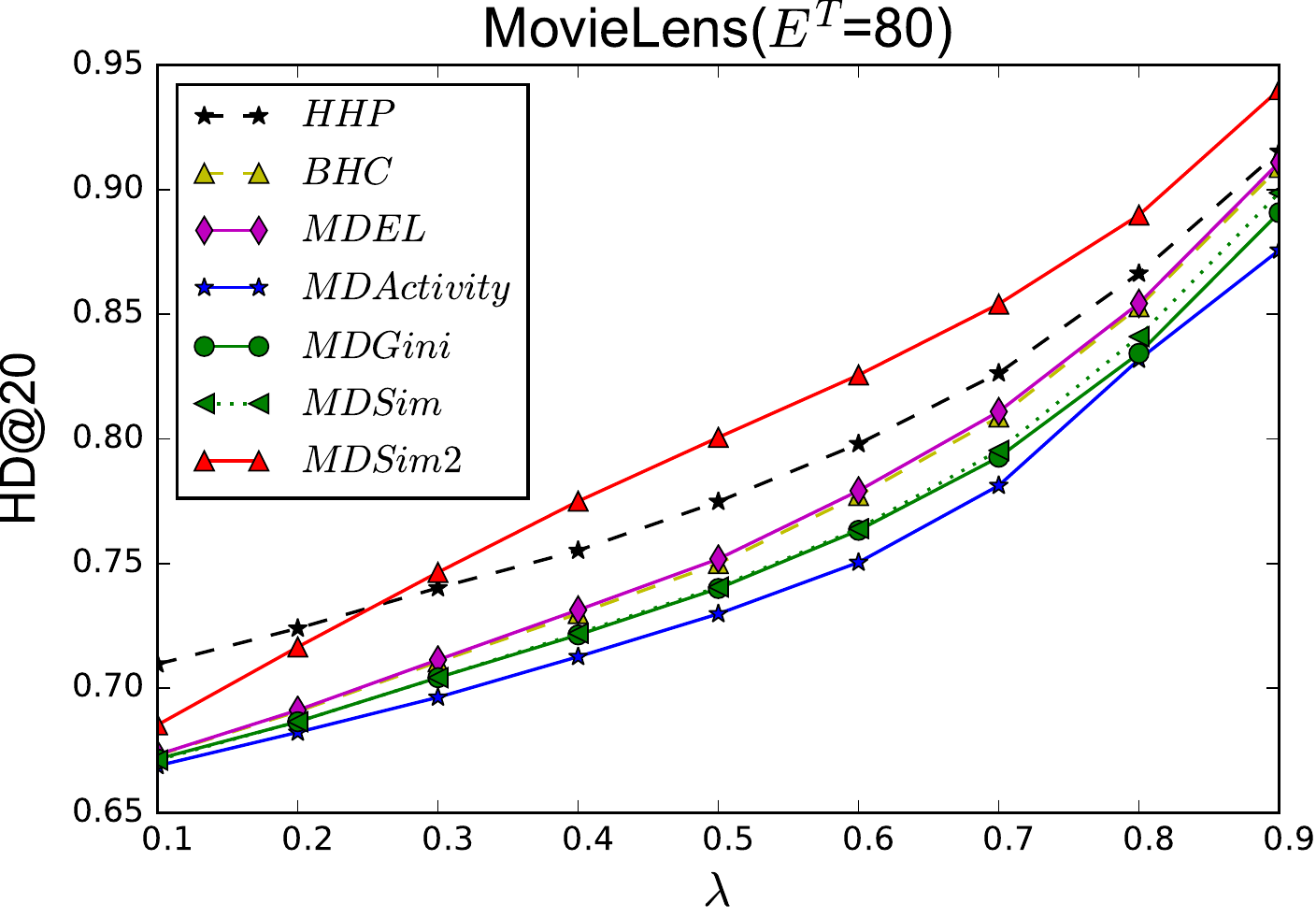}
		\includegraphics[width=0.32\textwidth]{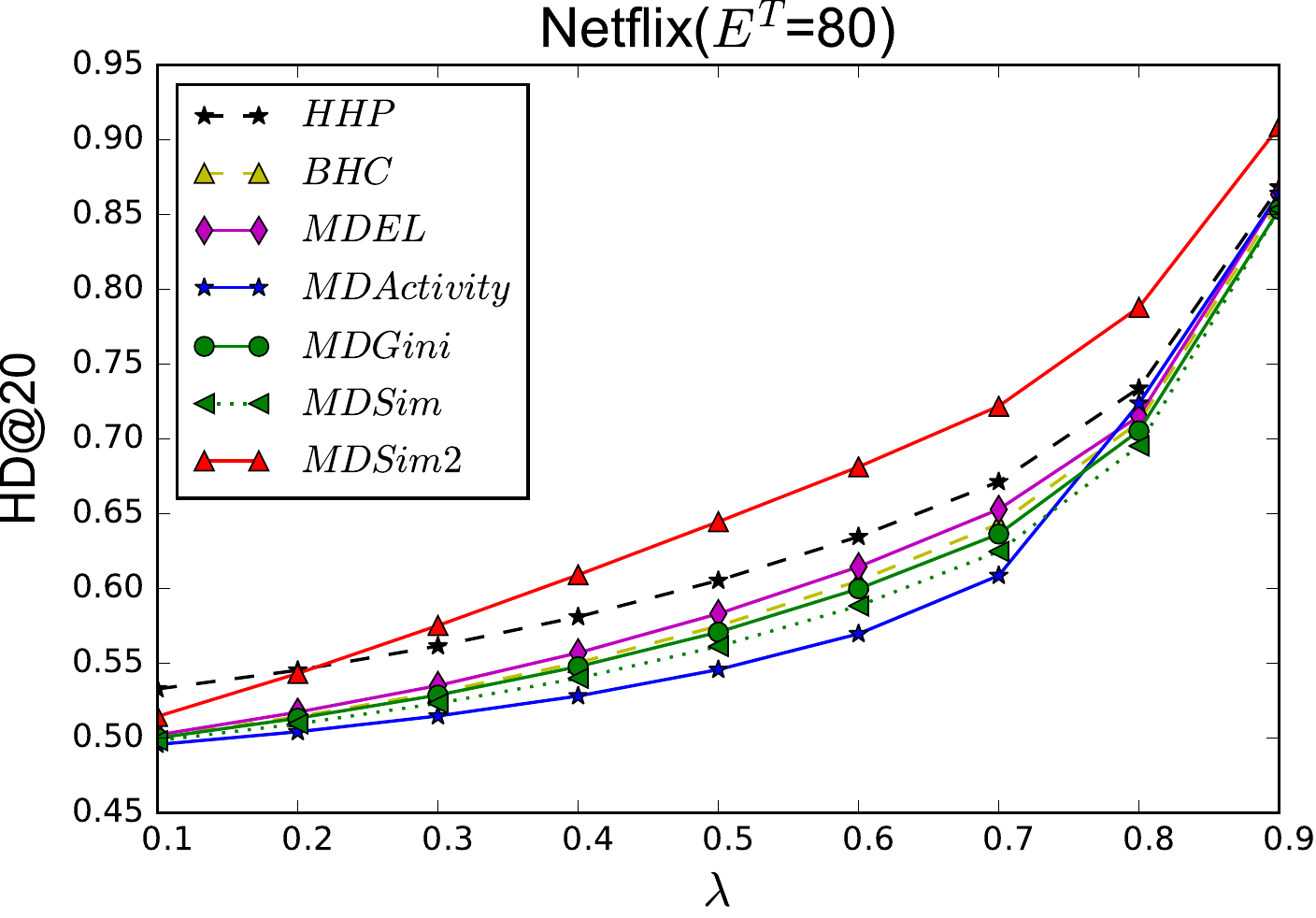}
		\includegraphics[width=0.32\textwidth]{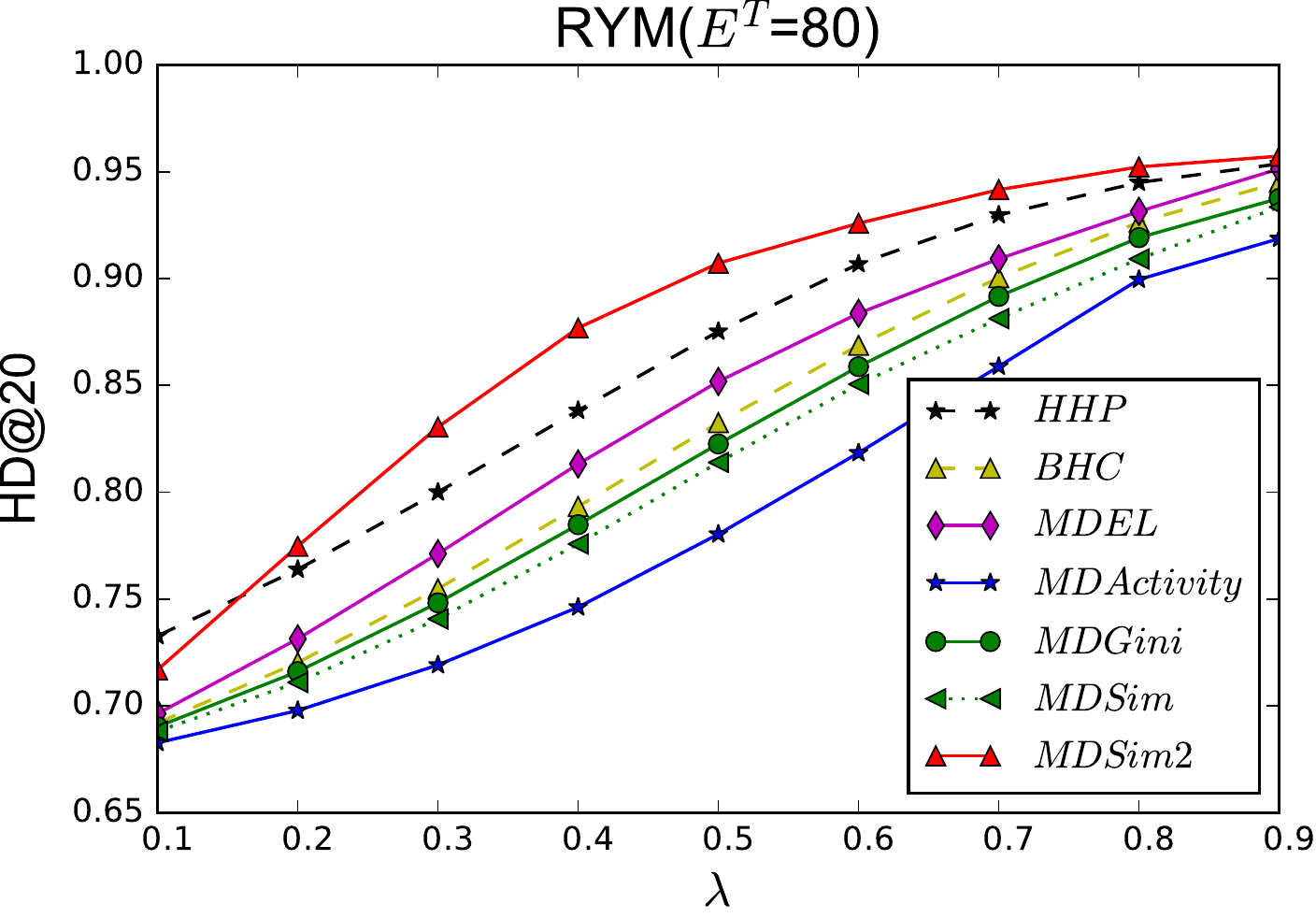}}
	
	\caption{The intra-diversity and inter-diversity of different models on three data sets, MovieLens($E^T$=80), Netflix($E^T$=80) and RYM($E^T$=80), respectively. The recommendation length is set to 20. Parameter $\lambda$ is traversed from $0.1$ to $0.9$, with the step of $0.1$.}
	\label{fig:overall-performance}
\end{figure*}

In Sec.III, we measure recommendation diversity as the total number of distinct items that are being recommended across all users, then one could possibly argue that, this kind of diversity could be easily improved by recommending more newly-released items. Thus, here we further evaluate the intra-diversity and inter-diversity of ExTrA-based models, measured by intraD-I@K and HD@K, respectively. The overall performance of diversity for all the methods on three datasets (MovieLens($E^T$=80), Netflix($E^T$=80), and RYM($E^T$=80)) are shown in Figure~ \ref{fig:overall-performance}.

From Table~\ref{tab:final results}, we inferred that, as $\lambda$ changed from $0.5$ to $0.9$, the accuracy followed the shape of first increasing, and then decreasing, and the diversity of all the proposed methods keep increasing. Apparently, the intra-diversity (intraD-I@20) and inter-diversity (HD@20) in Figure~\ref{fig:overall-performance} show the same trends, one except case is the MDSim2 on MovieLens for intra-diversity. Figure~\ref{fig:subfig:b} shows the comparison results of the proposed 5 ExTrA-based methods with the state-of-the-art methods HHP and BHC for intra-diversity, we could easily find out that on Netflix and RYM, MDSim2 model shows the best overall performance, which is followed by HHP. The MDEL, BHC, MDGini and MDSim models have almost the same performance and their performance are ranked after HHP. The MDActivity is the worst one but still much better than MD (see Sec.III). For the inter-diversity, the performance of the 7 methods on three datasets are consistent with the above discussions.

\subsection{Effect of Sparsity of data sets}

\begin{figure*}[!ht]
	\centering
	\subfigure[The sparse data sets.]{ 
		\label{fig:spr:a} 
		\includegraphics[width=0.32\textwidth]{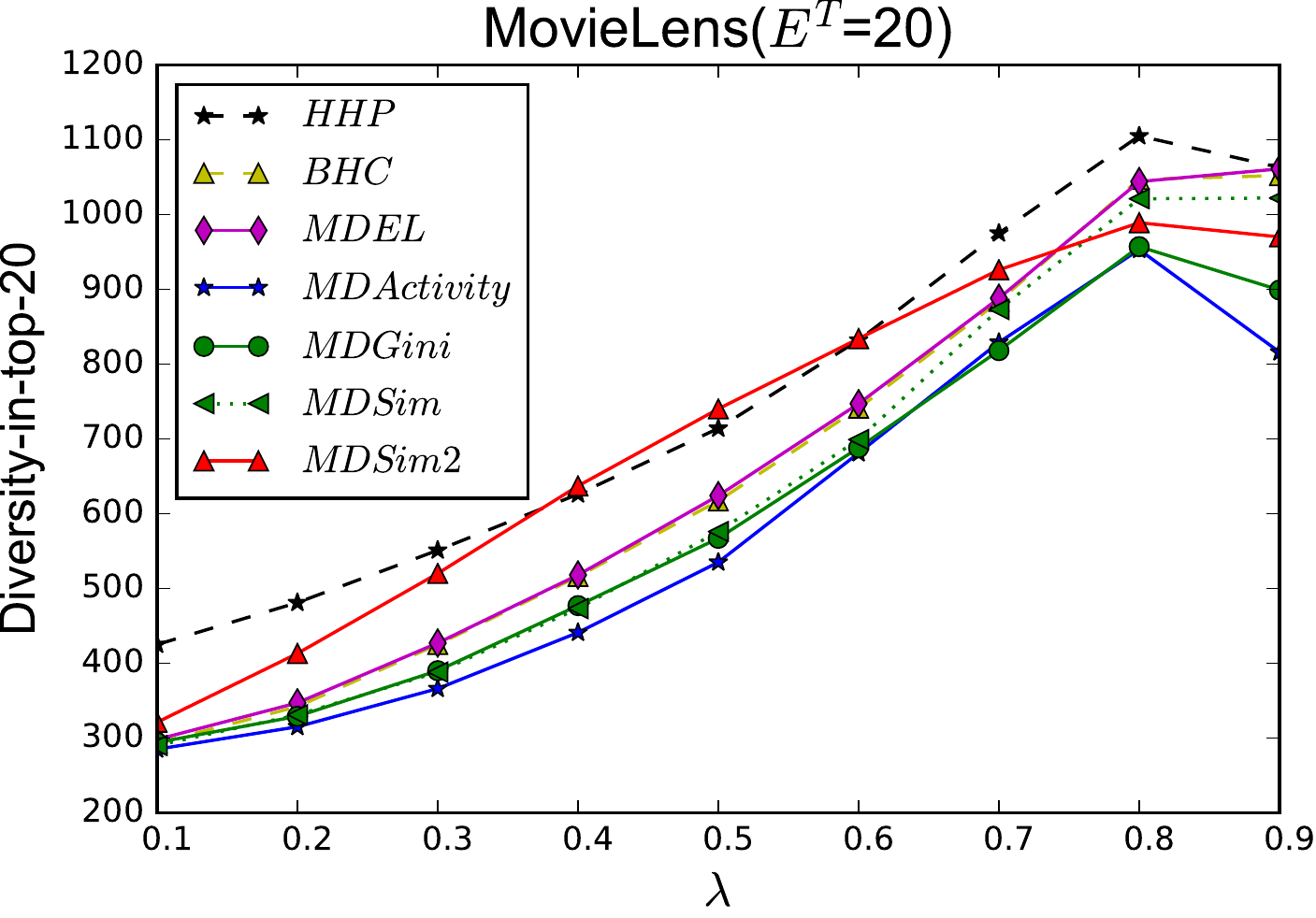}
		\includegraphics[width=0.32\textwidth]{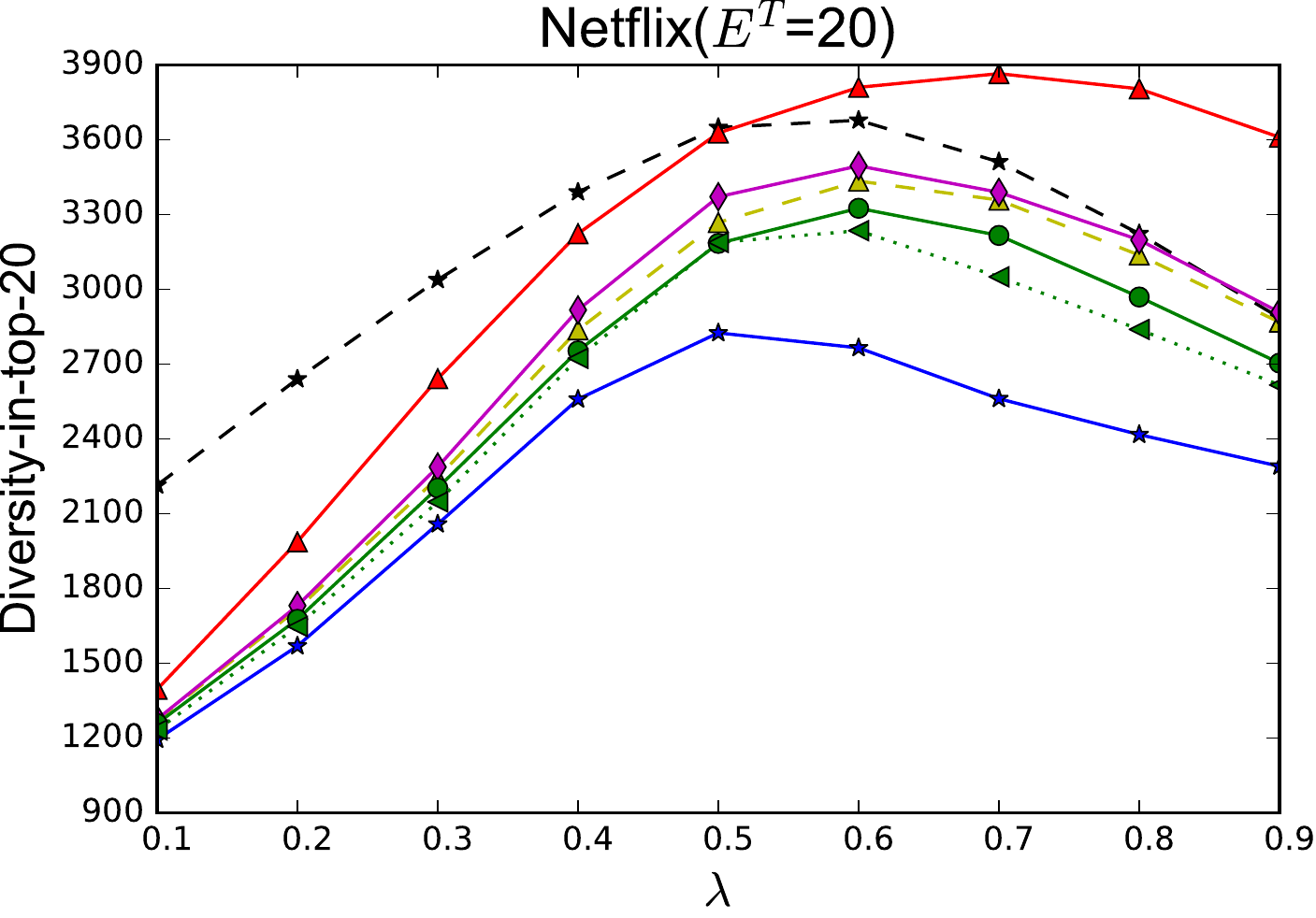}
		\includegraphics[width=0.32\textwidth]{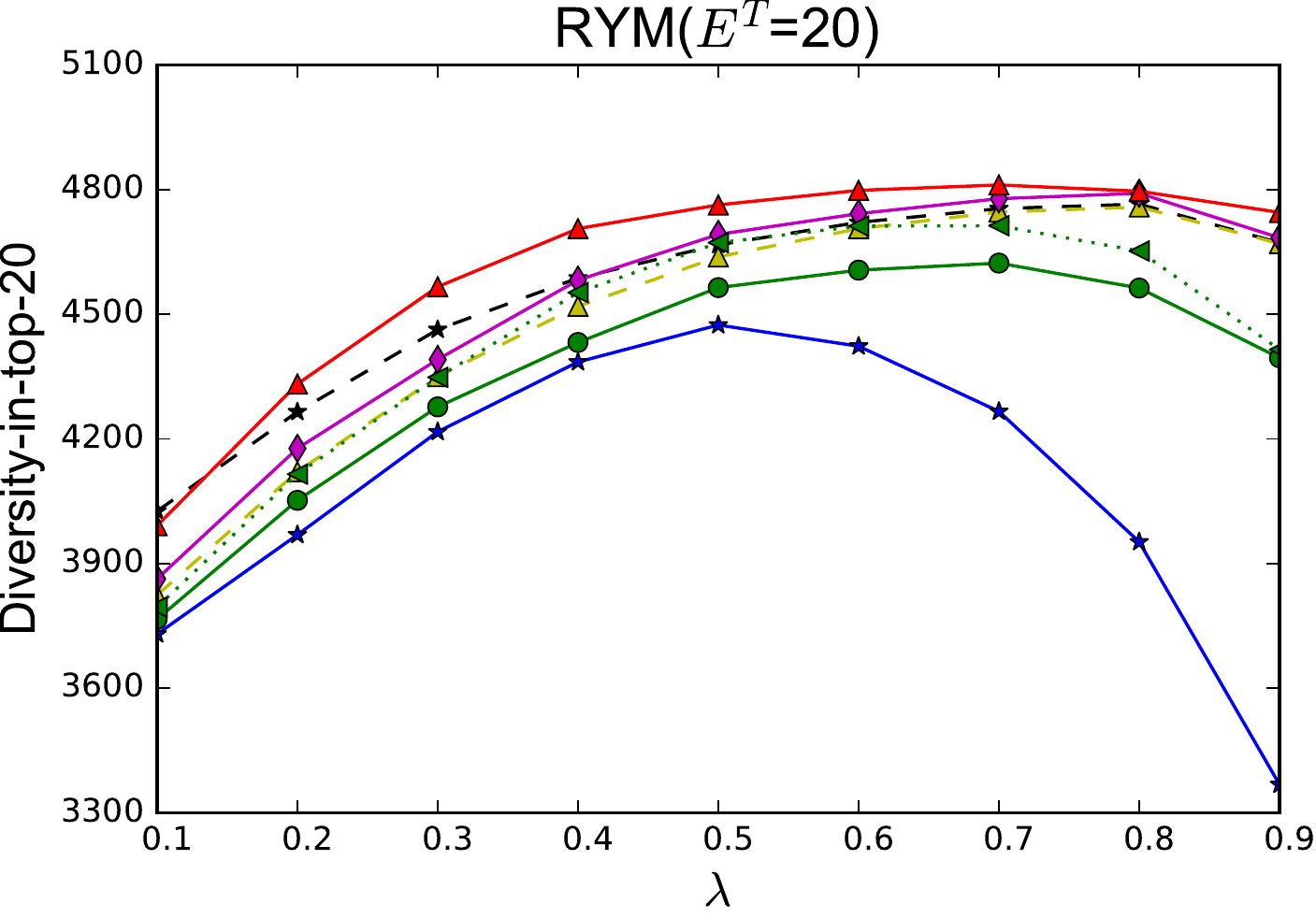}
		
	}	
	
	\vspace{5pt}
	\subfigure[The dense data sets.]{
		\label{fig:spr:b} \includegraphics[width=0.32\textwidth]{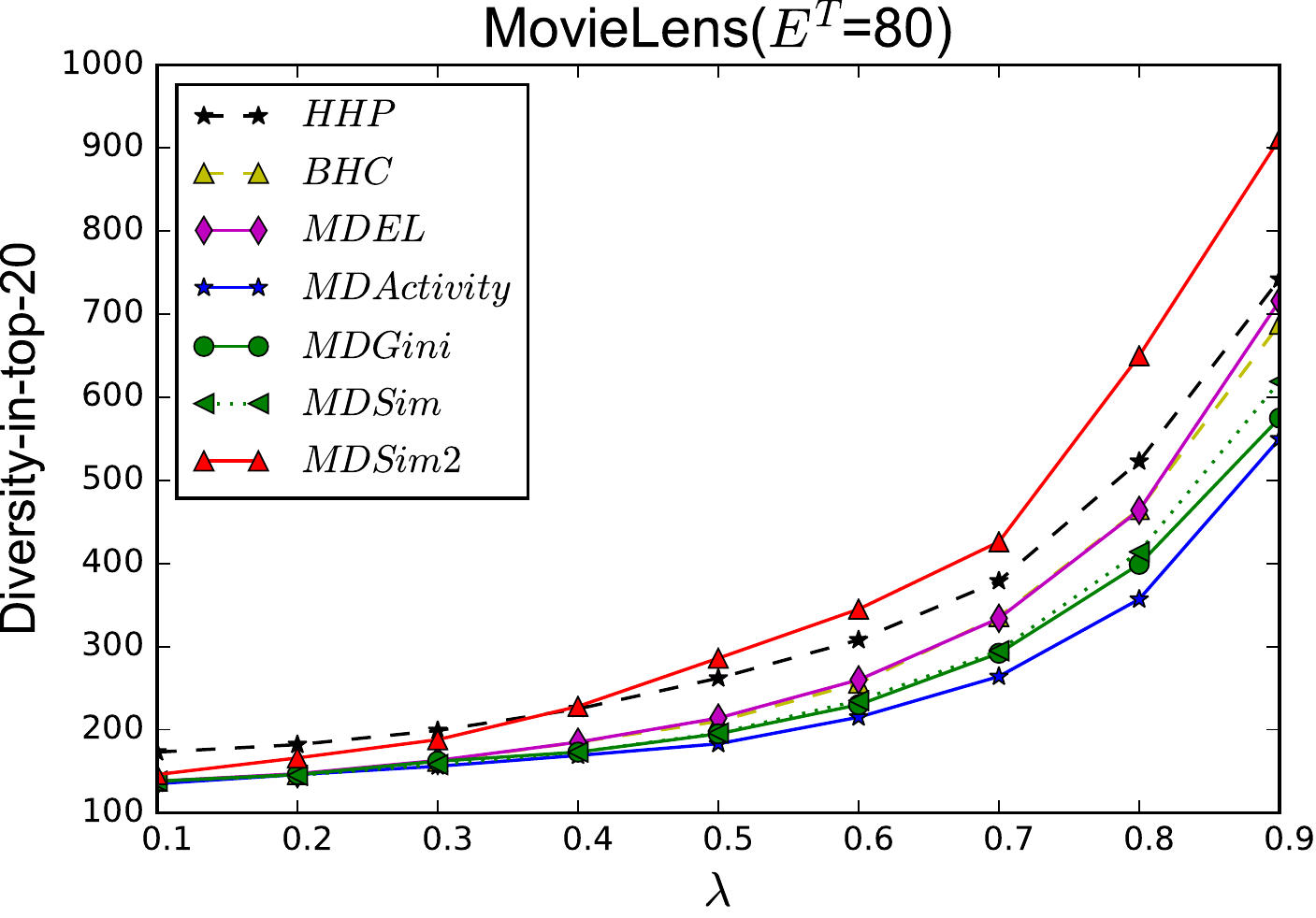}\includegraphics[width=0.32\textwidth]{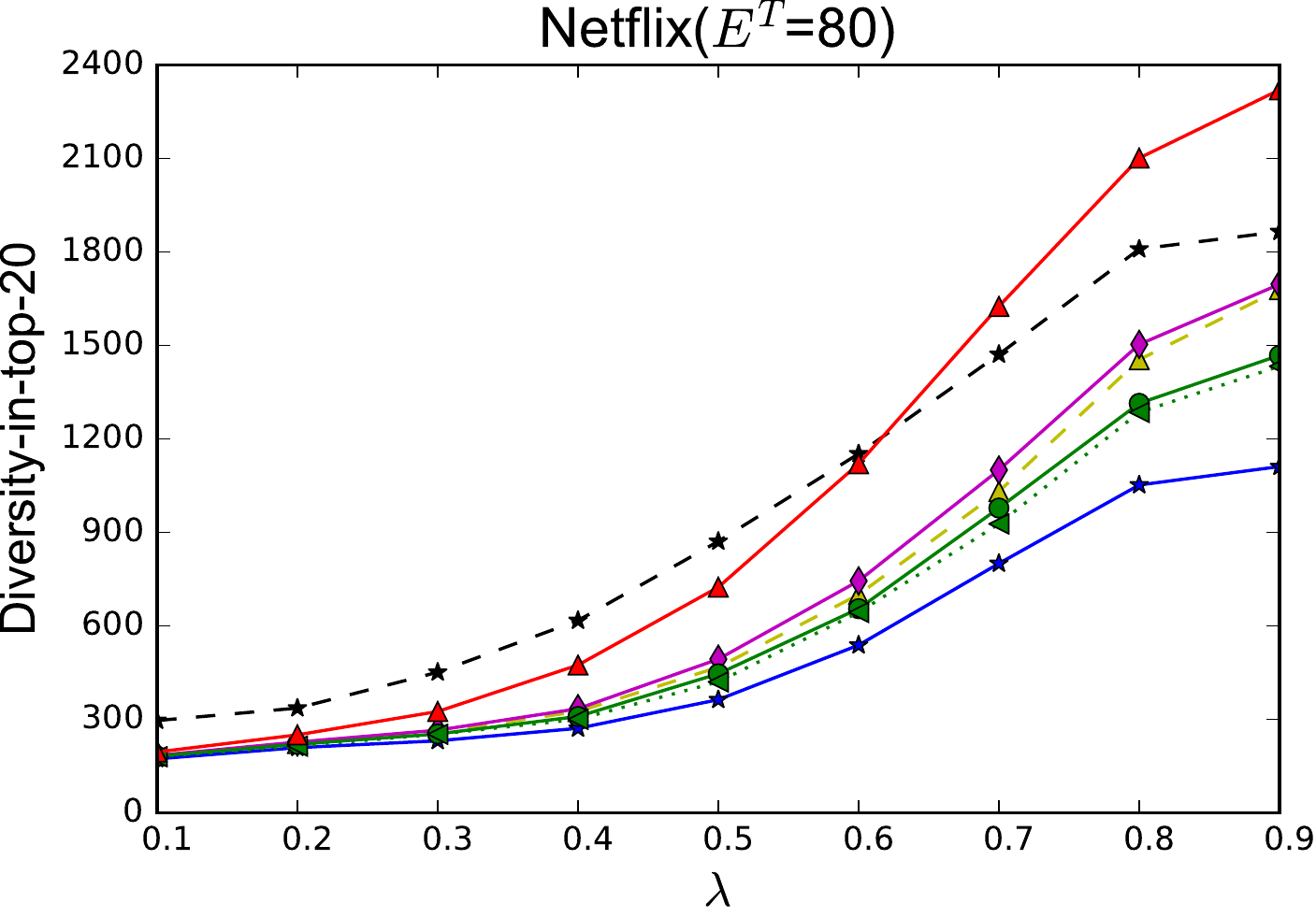}
		\includegraphics[width=0.32\textwidth]{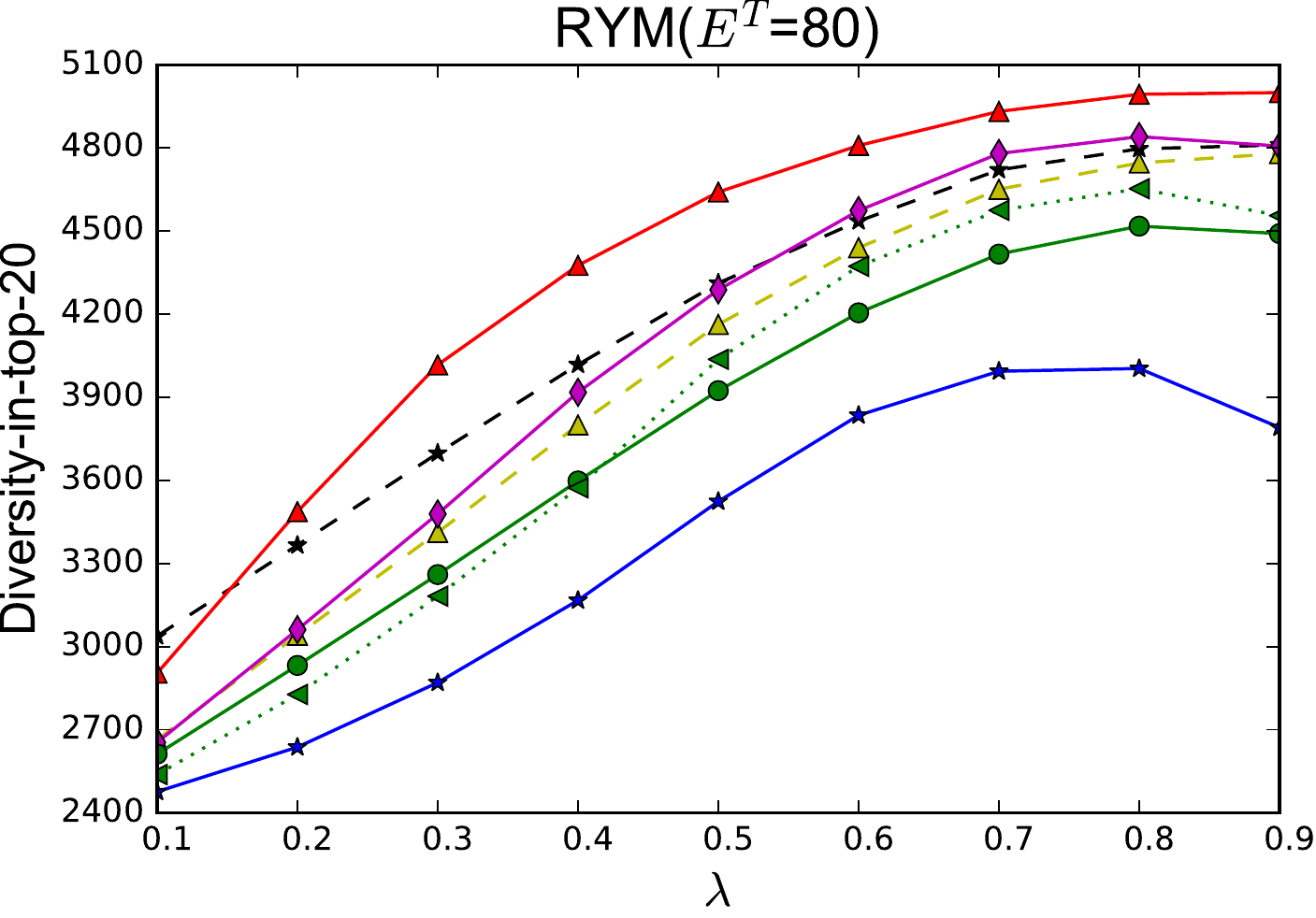}	
		
	}
	
	\caption{Diversity comparison of  different models on sparse and dense data sets, with the recommendation length of 20.} 
	\label{fig:sparsity}
\end{figure*}

Figure~\ref{fig:sparsity} shows the performance of proposed 5 ExTrA-based methods and 2 state-of-the-arts methods on data sets of different sparsity, $E^T$=20 and $E^T$=80. Firstly, we focus on data sets of $E^T$=80 (\ref{fig:spr:a}). For all 3 datasets, MDSim2 shows the best performance, HHP is a little lower than MDSim2, and the performance of BHC are all close to MDEL. But in the real applications, some recommendation systems often suffer the problem of cold start problem (which is simulated by sparse data in our case). Our proposed methods are based on extracting different features of users, thus, whether each of them could keep effective in the cold start scene? To answer this question, we test these methods on three datasets, which simulates the cold start problem: MovieLens($E^T$=20), Netflix($E^T$=20) and RYM($E^T$=20). The performance of these models on sparse datasets are shown in  Figure~\ref{fig:spr:a}. The performance of diversity for all the 7 methods on data sets of $E^T$=20 are similar with the performance on data sets of $E^T$=80, which confirms that our proposed methods are robust in the cold start condition.

\subsection{Effect of Recommendation Length}

\begin{figure*}[!ht]
	\centering
	\subfigure[The coverage diversity]{ 
		\label{fig:len:a} 
		\includegraphics[width=0.31\textwidth]{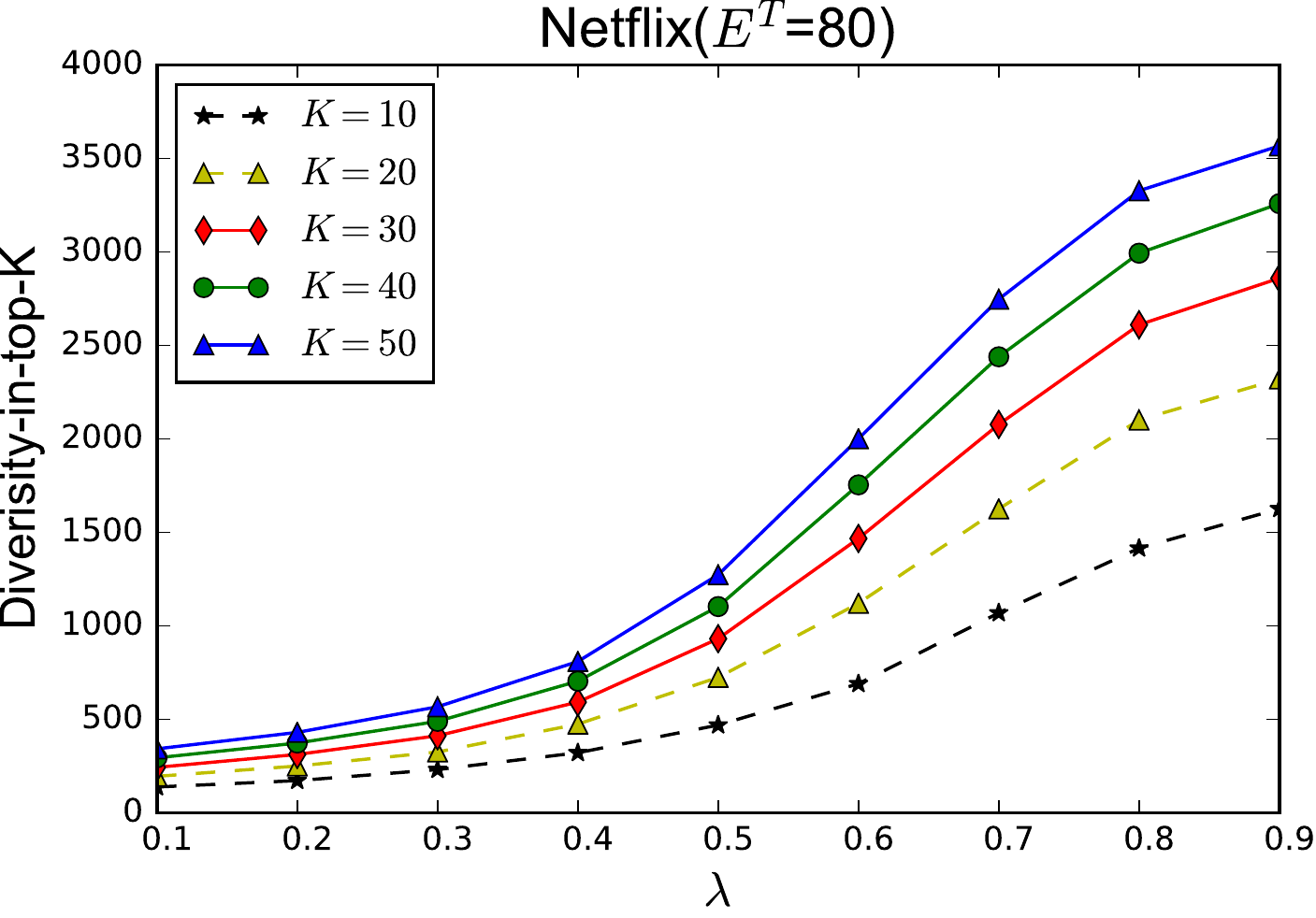}
	}	
	\subfigure[The intra-diversity]{
		\label{fig:len:c} 
		\includegraphics[width=0.31\textwidth]{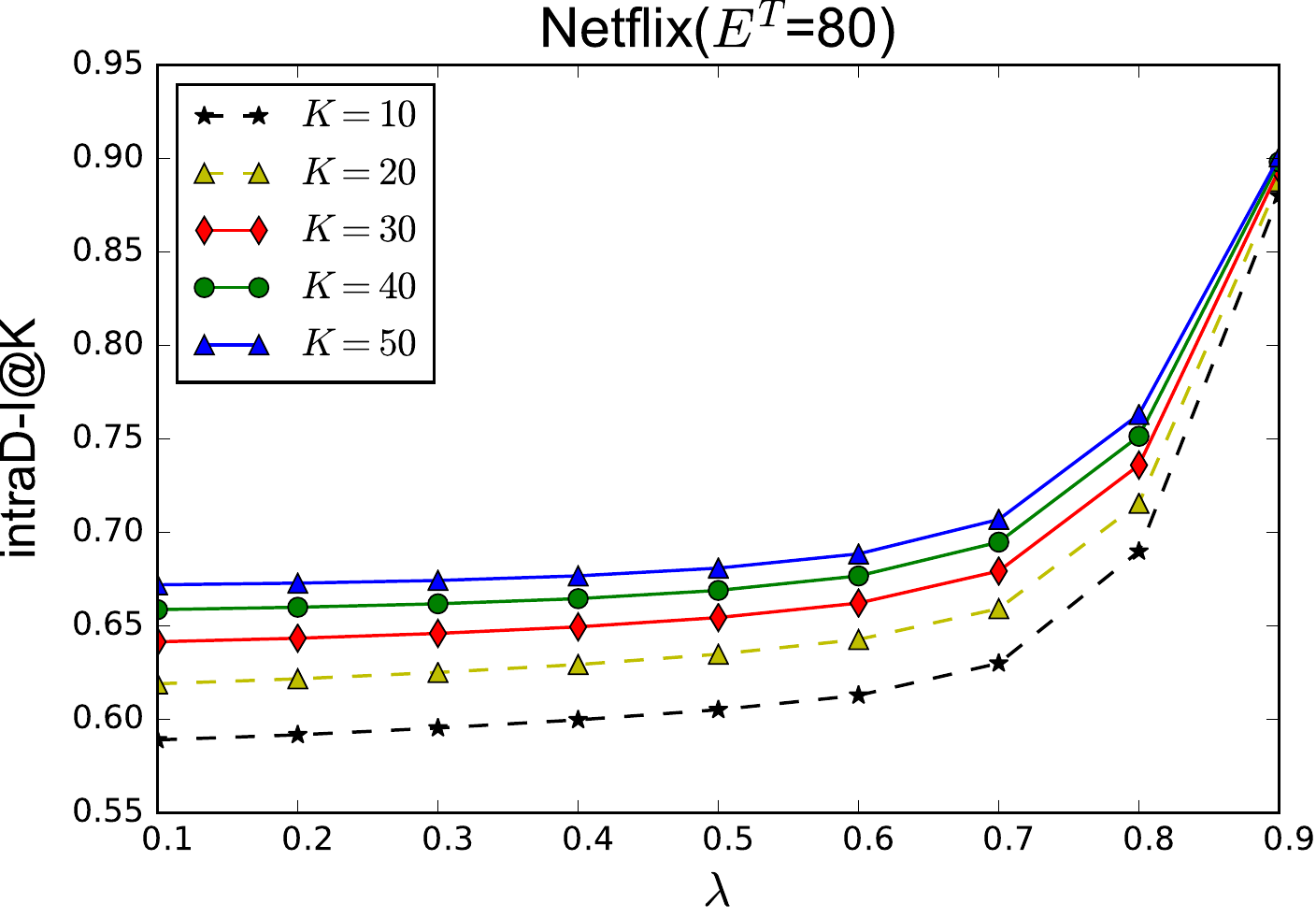}
	}
	\subfigure[The inter-diversity]{ 
		\label{fig:len:e} 
		\includegraphics[width=0.31\textwidth]{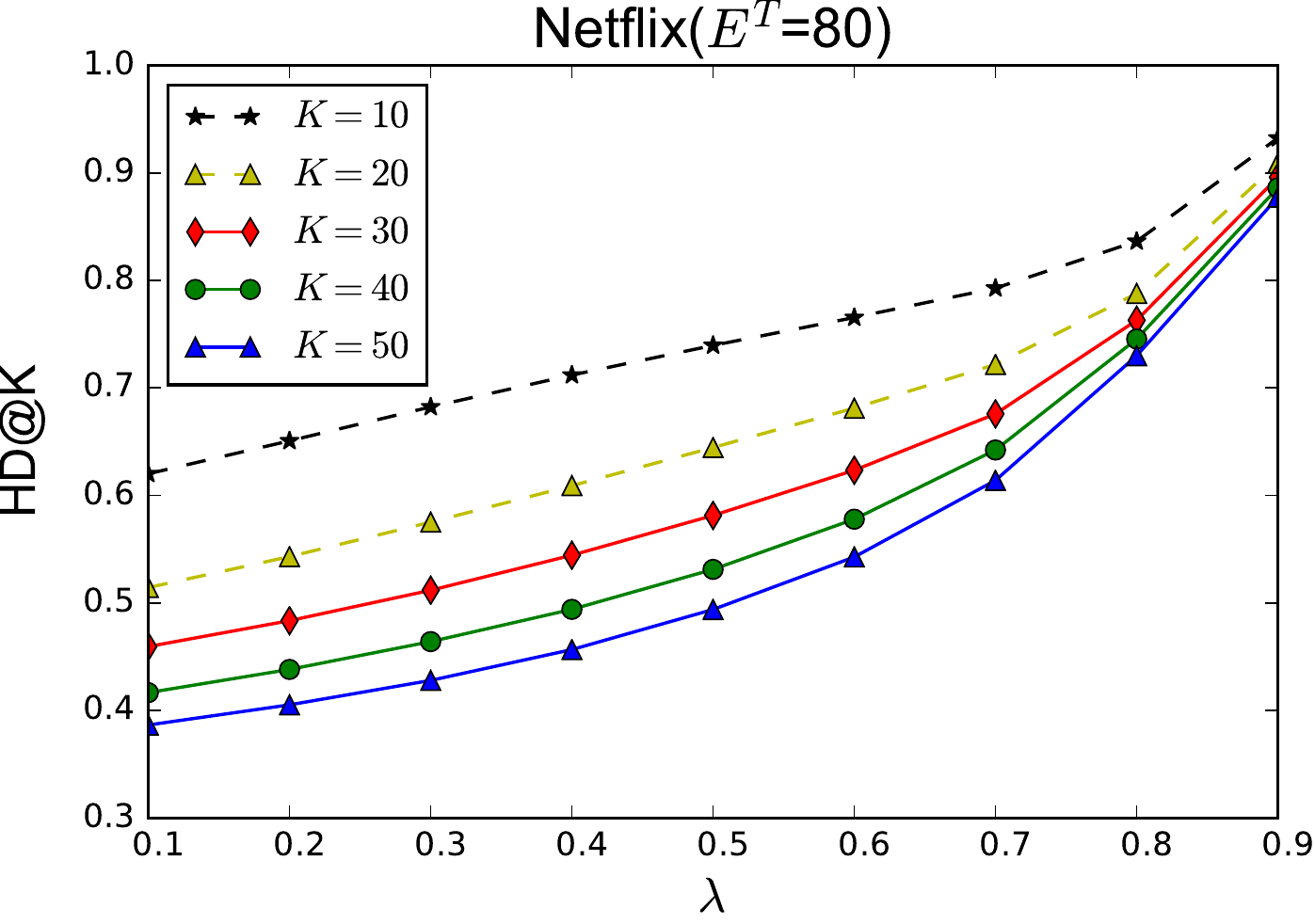}
	}
	\caption{The diversity of the best-performed MDSim2 model on the Netflix($E^T$=80) data set, with the recommendation length of 10, 20, 30, 40 and 50.} 
	\label{fig:rec-length} 
\end{figure*}

Figure~\ref{fig:rec-length} shows the performance comparison of the best-performed MDSim2 model on Diversity-in-top-K, intraD-I@K and HD@K on Netflix($E^T$=80) with different values of $K$. We observe that, with any recommendation length $K$, the tendency of performance are similar with the change of $\lambda$. In Figure~\ref{fig:len:a}, from left to right, Diversity-in-top-K for five values of $K$ have a similar shape: monotonically increasing; for intraD-I@K in Figure~\ref{fig:len:c}, the tendency is keeping increasing, but in particular, the performance gaps between intraD-I@10 and intraD-I@50 have narrowed from $\lambda$ = 0.7 to $\lambda$ = 0.8. It is not hard to understand this result with reference to Table~\ref{tab:final results}, as a large number of new items are added to the recommendation lists, from $1100$ to $1503$, it would affect the dis-similarity within each user's recommendation list. The results in Figure~\ref{fig:len:e} show that the longer is the recommendation length, the lower is the hamming distance. Therefore, when selecting recommendation length in real applications, it is not the truth that longer is better. It depends on what kind of measurement and the selection of $\lambda$.

\section{Relationship with some related Works}
\label{sec:related-works}
Users' satisfaction with recommendation results depends not only on prediction accuracy, but also on some other aspects of the recommendation quality such as diversity of the recommendation  lists. So far, some works have been conducted for the objective of increasing recommendation diversity, which could be divided into the intra-diversity and the inter-diversity (also known as aggregate diversity). The intra-diversity describes the diversity of the items in a user's recommendation list, thus, increasing the diversity means avoiding over specialization of items in a recommendation list. Strategies developed so far for increasing the intra-diversity mostly calculate the quality of an item based on its dissimilarity to the items that are already added into this user's recommendation list. The inter-diversity describes the dissimilarity between recommendation lists for each pair of users in the system. 

At present, there are many works on addressing how to improve the diversity of recommender systems in different application areas. Wu et al.~\cite{WuHao2014Oiar} introduced a simple yet elegant method to address this challenge from the aggregate perspective in folksonomy-based social systems. Belem et al. \cite{10.1145/2801130} considered three factors, the relevance, explicit topic diversity, and novelty conjointly in tag recommendations. Wu et al.~\cite{WuWen2018Prdb} took into account users' personality and proposed a generalized, dynamic personality-based greedy re-ranking approach to improve the personalized diversity in web applications. Yu et al.~\cite{YuTing2019RwdA} proposed an adaptive trust-aware recommendation model to improve the trade-off strategy of accuracy and diversity by studying the trust relationships among users, which could balance and adapt individual and aggregate diversity measures.

All these works utilize side information, more or less. However,  there are many constraints in the real application scenarios when utilizing side information. Thus, in the following, we will describe the efforts that increase the diversity of recommender systems by improving the exposure of niches in the diffusion-based model without introducing more side information. 
There are two lines of research that try to fulfill this task. 

The first line improves the diffusion process on the bipartite network particularly for the exposure of niche items~\cite{zhou2007bipartite,zhou2010solving,liu2011information,lu2011information,nie2015information,an2016diffusion}. The second line of research tries to extract core users from all users in the system~\cite{zeng2014uncovering,cao2016identifying} and rely on only these core users instead of all the users to generate recommendations. 

Zhou et al.~\cite{zhou2010solving} designed a nonlinear hybrid model of heat-spreading (HeatS, also known as HC) and ProbS (also known as MD), called Hybrid of HeatS and ProbS (HHP), which achieves significant improvements in both accuracy and diversity. Both of HC and MD work by assigning collected items of the target user an initial level of "resource", and then redistribute it via a transformation function from an item to another item via common users. The  recommendation list  is obtained by sorting the uncollected items according to the obtained resource in descending order. The difference between MD and HC is that, the niche items that to be recommended in HC would actively absorb more resource from common users than the niches that passively receive averagely allocated resource in MD. In this way, niche items are pushed to the head of recommendation lists and very popular items are rejected in HC. In MD, the popular items are generally assigned more resource. As a result, the candidate items in HC are mostly niches, which leads to high diversity but very low accuracy and on the contrary, MD with high accuracy but low diversity. By a non-linear hybrid of HC and MD, HHP balances the resource distribution during the process of resource assignment and improve the diversity without losing accuracy. Another effective method modified delicately from original HC, named Biased Heat Conduction (BHC)~\cite{liu2011information}, also makes a good trade-off on accuracy and diversity. The recommendation procedure for BHC is the same with HC, but in the second step from  users to candidate items, the resource absorbed by niche candidates are decreased by a manually tuned parameter. Further statistical analysis on the recommendation lists in this paper show that the items with large or small degrees are all recommended  frequently in BHC, but large-degree items are recommended more frequently in MD and small degree items are recommended more frequently in HC. It suggests that BHC could simultaneously identify the public and personalized tastes of users, resulting in better performance than the standard HC algorithm.

The other line of research on bipartite graph tries to extract core users from the system. Zeng et al.~\cite{zeng2014uncovering} found that in each online system there exists a group of core users who carry more useful information for recommendation. They designed core user extraction methods in the individual level and the system level respectively that could enable the recommender systems to achieve $90\%$ of the accuracy of the standard procedure by utilizing only $20\%$ of the users to generate recommendations. In practical applications, the most time-consuming process for this work is to extract core users, which could be calculated offline that enables the online recommendation process efficient. Cao et al.~\cite{cao2016identifying} proposed to identify core users based on trust relationships and interest similarity to acquire more accurate recommendations. In this work, the trust and interest similarity between all user pairs are calculated and sorted first, and two strategies based on frequency and weight of location are used to select core users.  One method is called frequency-based strategy, namely, to select users who appear the largest number of times in all other users' nearest neighbor list. The other one is rank-based strategy, which selects users who have the highest weight of location in all other users' nearest neighbor list. The results show that core users usually appear in many users' top-K neighbor lists with small ranking numbers.  They got the similar conclusion with Zeng's study that the core users usually carry more useful information for recommendation, and the RSes can make use of only core users to achieve satisfactory recommendation accuracy. 

Our proposed ExTrA-based methods try to enhance the role of fabricated experts in discovering niche items and thus fall into the latter research line mentioned above. In contrast to the other approaches of this line, however, it does not require any semantic metadata (which is often not available or incomplete) but calculates the expertise of a user based on the history data. Also, it utilizes all the users instead of only core users, but highlights the weights of core users compared with other users. Thus, it is not a hybrid but a new diffusion-based approach. The first research line reconstructs the network based on the diffusion characteristics to push niche items from the long tail to the head to improve the diversity of the RSes, which is in line with our purpose. However, we put our emphasis on the roles of user nodes in the bipartite graph rather than the edges, which means that when using the ExTrA-based methods to calculate the predicted preferences for a user-item pair, these reconstructed approaches (HHP, BHC et al.) can also benefit from it.

\section{Conclusions and Future Works}
\label{sec:conclusion}

In this paper, we introduce a family of approaches to extract fabricated experts from all users in recommender systems, and highlight them in the mass diffusion model. Comprehensive empirical experiments witness the significant diversity improvement brought by the proposed methods, with no or trivial accuracy loss of recommendation results. Note that, some delicately designed expert discovering methods might obtain better performance than our proposed ones, however, our motivation is not proposing the best expert extraction approach for more accurate and diverse predictions, but aiming at highlighting neighbor users' different capability of recommending relevant and personalized items to the target user.

This work might shed light on several interesting directions for the future research. First, additional expert selection criteria should be explored for the given application domains. This may introduce more side information and also more sophisticated techniques (for example, knowledge graph-based methods~\cite{WangHongwei2018RPUP}) depending on the specific applications, which comes with possibly significant increase in computational complexity.
Second, although the MD model is a special case of memory-based Collaborative Filtering (CF) with the RA similarity~\cite{YuFei2016NraA}, to explore the limitations of the proposed methods, the usefulness of highlighting the fabricated experts should be checked for the model-based CF (such as the matrix factorization models), and the memory-based CF models with common similarity measures, such as the Cosine similarity and the Jaccard similarity. 
Third, improvement of recommendation diversity when recommending for a group of users (instead of individual users)~\cite{CaoDa2018AGR} also constitutes interesting topics for the future research.

\section{Acknowledgements}
This research is funded by UESTC Fundamental Research Funds for the Central Universities under Grant No.: ZYGX2016J196.


\bibliographystyle{ieeetr}
\bibliography{mybibfile}

\end{document}